\documentclass{aa}  

\usepackage{graphicx}\usepackage{rotating}

\usepackage{multirow}

\usepackage{epsfig}
\usepackage{subfigure}

\usepackage{graphicx}
\DeclareGraphicsExtensions{.ps}
\usepackage{rotating}

\usepackage{caption}
\usepackage{graphicx}
\usepackage{lscape}
\usepackage{amsmath}
\usepackage{amssymb}
\usepackage{amsfonts}
\usepackage{tabularx}
\usepackage{natbib}
\usepackage{graphicx}
\usepackage{epsfig}
\usepackage{verbatim}
\usepackage[flushleft]{threeparttable}
\include{journaldefs}
\usepackage{url}
\usepackage{textcomp}
\usepackage{float}
\usepackage[T1]{fontenc} 
\usepackage{aecompl}
\usepackage{adjustbox}
\usepackage{txfonts}
\begin{document}

   \title{Water megamaser emission in hard X-ray selected AGN}
   %\footnote{Table 1A is only available in electronic form at the CDS via anonymous ftp to cdsarc.u-strasbg.fr (130.79.128.5) or via http://cdsweb.u-strasbg.fr/cgi-bin/qcat?J/A+A/}

   \author{F. Panessa
          \inst{1}
          \and
          P. Castangia\inst{2}
            \and
           A. Malizia\inst{3}
          \and
          L. Bassani\inst{3}
          \and
          A. Tarchi\inst{2}
          \and
          A. Bazzano\inst{1}
          \and
          P. Ubertini\inst{1}
          }

   \institute{Istituto di Astrofisica e Planetologia Spaziali di Roma (IAPS-INAF), Via del Fosso del Cavaliere 100, 00133 Roma, Italy\\
              \email{francesca.panessa@inaf.it}
         \and
             Osservatorio Astronomico di Cagliari (OAC-INAF),Via della Scienza 5,  09047 Selargius (CA), Italy 
        \and     
             Osservatorio di Astrofisica e Scienza dello Spazio di Bologna (OAS-INAF), Via P. Gobetti 101, 40129 Bologna, Italy
            }

   \date{Received 23 December 2019 / Accepted 12 June 2020}

% \abstract{}{}{}{}{} 
% 5 {} token are mandatory
 
  \abstract
   {Water megamaser emission at 22\,GHz has proven to be a powerful tool for astrophysical studies of active galactic nuclei (AGN) because it allows an accurate determination of the mass of the central black hole and of the accretion disc geometry and dynamics. However, after searches among thousands of galaxies, only $\text{about}$ 200 of them have shown such spectroscopic features, most of them of uncertain classification. In addition, the physical and geometrical conditions under which a maser activates are still unknown.}
   {We characterize the occurrence of water maser emission in an unbiased sample of AGN by investigating the relation with the X-ray properties and the possible favourable geometry that is required to detect water maser.}
   {We searched for 22\,GHz maser emission in a hard X-ray selected sample of AGN, taken from the INTEGRAL/IBIS survey above 20 keV. Only half of the 380 sources in the sample have water maser data. We also considered a volume-limited sub-sample of 87 sources, for which we obtained new observations with the Green Bank and Effelsberg telescopes (for 35 sources). We detected one new maser and increased its radio coverage to 75\%.}
   {The detection rate of water maser emission in the total sample is 15$\pm$3\%. This fraction increases to 19$\pm$5\% for the complete sub-sample, especially when we consider type 2 (22$\pm$5\% and 31$\pm$10\% for the total and complete samples, respectively) and Compton-thick AGN (56$\pm$18\% and 50$\pm$35\% for the total and complete samples, respectively).
   No correlation is found between water maser and X-ray luminosity. We note that all types of masers (disc and jet) are associated with hard X-ray selected AGN.}
   {These results demonstrate that the hard X--ray selection may significantly enhance the maser detection efficiency over comparably large optical or infrared surveys. A possible decline in detection fraction with increasing luminosity might suggest that an extremely luminous nuclear environment does not favour maser emission. The large fraction of CT AGN with water maser emission could be explained in terms of geometrical effects. The maser medium would then be the very edge-on portion of the obscuring medium.}


   \maketitle
%
%-------------------------------------------------------------------

\section{Introduction}

One of the most common maser emission lines arises from the water rotational transition levels 6$_{16}$ and 5$_{23}$, which emit at 22\,GHz in the radio domain. Extra-galactic water masers trace warm (T$_{kin}$ > 300 K) and dense (10$^{7}$ cm$^{-3}$ $<$ n(H$_{2}$) $<$ 10$^{11}$ cm$^{-3}$) gas \citep{elitzur92,neufeld}.
Water maser sources with an isotropic luminosity below 10 L$_{\odot}$ are defined as kilo-masers, while at higher luminosity, they are defined as mega-masers. The latter are generally associated with the activity of active galactic nuclei (AGN), while kilo-masers are more commonly related to star formation in the host galaxy (this distinction should, however, be used with caution, see Sect. 4.2 in Tarchi et al. (\citeyear{tarchi11a}) and this work Sect.3).

The activity of water maser emission in AGN has been associated with three main different phenomena (see e.g. Lo \citeyear{lo05}; Tarchi \citeyear{tarchi12}). A typical triple-peak system of lines is associated with an accretion disc emission (one systemic, one blueshifted, and one redshifted), whose geometry and rotation velocities  can be traced by very long baseline interferometry (VLBI, e.g. Miyoshi et al. \citeyear{miyoshi}; Greenhill et al. \citeyear{green03a}).  
On the other hand, the interaction between the radio jet and the molecular clouds or the overlap along the line of sight between the molecular cloud and radio continuum emission from the jet might produce water maser emission in the form of a single broad redshifted (or blueshifted) line (e.g. Gallimore et al. \citeyear{gallimore01,gallimore96}; Henkel et al. \citeyear{henkel05}). Jet velocity and density were estimated in a reverberation mapping analysis \citep{peck}. 
Finally, in the case of the Circinus galaxy, the water maser emission has shown two different dynamic components in VLBI mapping: one associated with a warped disc, and another with a wide-angle nuclear outflow \citep{green03a}. 
Outflowing maser components have also been detected in NGC~3079 \citep{kondratko05}.

So far, more than 4000 galaxies have been searched for water maser emission, and detections have been obtained in about$\text{}$ 160 of them (180 when starburst galaxies are also included; Megamaser Cosmology Project, MCP\footnote{\url{https://safe.nrao.edu/wiki/bin/view/Main/MegamaserCosmologyProject}}), 
the majority are radio-quiet AGN in the local Universe (z$\le$0.05), classified as Seyfert 2 or low-ionisation nuclear emission-line regions (LINERs).
The overall detection rate in large maser surveys is rather low (e.g. Braatz et al. \citeyear{braatz97}; Greenhill et al. \citeyear{green03a}; Van den Bosch et al. \citeyear{van})
and is about 3\% for sources observed within the MCP, which mainly targets galaxies selected from large optical surveys, such as SDSS, 6dF, and 2MRS (Braatz et al. \citeyear{braatz15}; see also Greenhill et al. \citeyear{green03b}; Zhu et al. \citeyear{zhu}). 
Hagiwara et al. (\citeyear{hagiwara02,hagiwara03}) selected their targets based on the ratio of radio continuum to IR (60 $\mu$ and 100 $\mu$) flux densities from IRAS galaxies and obtained a slightly higher detection rate of 8\%.  \citet{henkel05} later confirmed that the far-IR selection favours maser detection. These authors found a detection rate of 22\% in a sample of northern galaxies with an IRAS point source flux density at 100 micron greater than 50 mJy. More recently, \citet{kuo18} found that galaxies with water maser detection tend to be associated with strong IR emission, as observed by the WISE telescope, thus offering a way to boost the detection rate to 6-15$\%$. It has also been suggested that radio emission is a suitable indicator for water maser emission \citep{zhang12,zhang17}, and the radio luminosities of maser galaxies indeed tend to be higher by a factor of 2–3 than those in non-masing galaxies \citep{liu}.

Finally, the fraction of water maser detection has been found to be around 26\% in a sample of Seyfert galaxies located within 20 Mpc. This suggests that an observational bias in terms of distance is also likely \citep{pg}.

The maser detection efficiency might be improved if high-luminosity objects were selected \citep{zhu}.
However, the largest fraction of nuclear water masers seems to be associated with type 2 Seyfert galaxies and a high level of X--ray obscuration \citep{green08}, in particular to Compton-thick AGN (CT AGN are defined as sources with an X-ray obscuration of N$_{H}$ $>$ 10$^{24}$ cm$^{-2}$. This is the inverse of the Thomson cross section), \citep{green03b, cast19}. This is in line with the predictions of unified models for AGN \citep{antonucci}, in which an obscuring torus that is aligned with the accretion disc at larger scales is responsible for the observed obscuration and for the optical classification of the AGN (see Padovani et al. \citeyear{padovani} for a review). Interestingly, the fraction of CT obscuration increases in disc masers \citep{green08}, as was confirmed by X-ray studies of known disc masers \citep{cast13, masini16}. In order to be detected, maser discs should be observed nearly edge-on to the observer line of sight. This suggest that the X--ray obscuring material and the maser disc are connected. Masers might indeed trace molecular material associated with the torus or the outer regions of the accretion disc. All the different proposed geometries \citep{elitzur06, tilak, masini16} take into account that long path lengths are needed to produce maser amplification, therefore the observer line of sight has to be close to an edge-on orientation. In this respect, warped discs, as have been observed in the prototype NGC~4258 \citep{herrnstein}, increase the probability that the line of sight is intercepted. The disc has to be warped to be directly illuminated by the X-ray radiation from the central engine, as envisaged by the theory of maser production \citep{neufeld}.
More recently, \citet{darling} discussed the interesting possibility that some water maser sources might be detected that are associated
with inclined accretion discs (more than 10 degrees from edge-on) and orbit massive black holes. The detection would be made by the lensing or deflection of in-going systemic maser features.

Even though objects with higher X-ray luminosity and/or higher column density more likely host masers, there is no large sample of AGN with X-ray data available so far for a target selection, nor has a similar type of study been performed on a statistically meaningful basis.
This work aims at filling this gap and at providing some useful means to improve the maser detection efficiency by pre-selecting targets from hard X-ray surveys, which so far are the least biased in terms of intrinsic AGN absorption. 

Throughout this paper we assume a flat $\Lambda$ cold dark matter cosmology with ($\Omega_{\rm M}$, $\Omega_{\rm\Lambda}) = (0.3$,0.7) and a Hubble constant of 70 km s$^{-1}$ Mpc$^{-1}$ \citep{jarosik}. 

%--------------------------------------------------------------------
\section{Sample definition}

We concentrate on a sample of active galaxies selected in hard X-rays (or soft gamma-ray band, defined above 20 keV). This waveband provides a very efficient way to find nearby AGN (unabsorbed and absorbed) because it is transparent to obscured regions or objects, that is, those that might be missed at other frequencies such as optical, UV, and even X-rays below 10 keV. Since 2002, the hard X--ray sky is being surveyed by INTEGRAL/IBIS \citep{ubertini} and subsequently by Swift/BAT \citep{gehrels04} at energies greater than $\sim$ 20 keV; various all-sky catalogues have been released based on the data collected by these two satellites (see e.g. Bird et al. \citeyear{bird16}; Baumgartner et al. \citeyear{baum} and Oh et al. \citeyear{oh}). These catalogues contain large fractions of active galaxies, 
$\text{about}$ 40\% of the INTEGRAL/IBIS and up to 70\% of the Swift/BAT sources. These two samples together provide the most extensive list of hard X-ray selected active galaxies known to date. 

For the purpose of this work, we used the large sample of AGN extracted from INTEGRAL/IBIS data. For comparison purposes alone, we consulted two samples extracted from Swift/BAT surveys (the 9-month and 70-month samples).

For INTEGRAL, we considered the sample of 272 AGN discussed by \citet{malizia12},
to which we added 108 sources that have been discovered or identified with active galaxies afterwards \citep{malizia16}. This set of 380 hard X-ray selected AGN represents our reference catalogue and was used as the main input for this work. The main advantage of this sample is that it is fully characterised in terms
of optical class, redshift, and X-ray properties, including information on the X-ray and hard X-ray fluxes and X-ray column density. Unfortunately, due to the INTEGRAL observing strategy, this sample is not complete or uniform, and to overcome this limitation, we considered a subset of AGN (all included in the sample of 380 objects) that instead represent a complete sample. This sample, which  
is fully discussed in \citet{malizia09}, consists of 87 galaxies that are detected in the 20--40 keV band and listed in the thirrd IBIS survey \citep{bird07}. We note that one source, IGR~J03184-0014, is not considered here as it was never again detected in subsequent INTEGRAL surveys.
To investigate maser emission in the entire sample of 380 AGN, we consulted the catalogues maintained on the website of the MCP, which is the largest and most comprehensive catalogue of all galaxies that have been surveyed for water maser emission at 22\,GHz \citep{reid,braatz10}. The catalogue has been updated on a regular basis to include all of the new observations
and associated findings. To integrate the MCP data and to cover our sample as far as possible, we also searched the literature for reports of water maser observations or detections. Finally, 35 galaxies belonging to the complete sample were observed for the first time in search for 22\,GHz water maser emission using the Effelsberg telescope and the Green Bank Telescope (GBT). We discovered new maser (Sect. 3). Table 1A lists all 380 INTEGRAL/IBIS AGN (see Appendix A for a detailed description).

\begin{figure*}
\begin{center}
\parbox{18cm}{
\includegraphics[width=0.45\textwidth,height=0.45\textheight,angle=0]{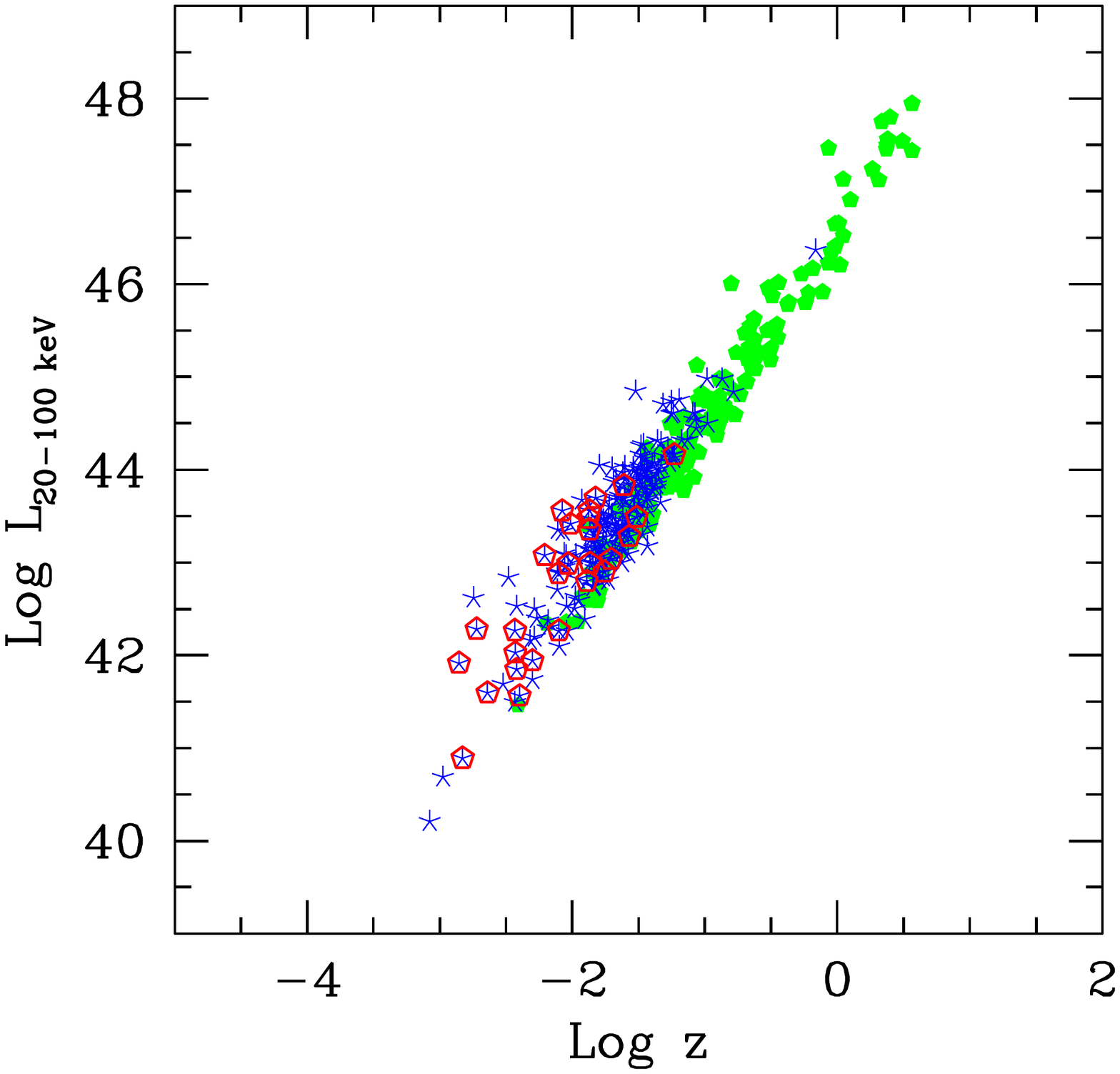}
\includegraphics[width=0.45\textwidth,height=0.45\textheight,angle=0]{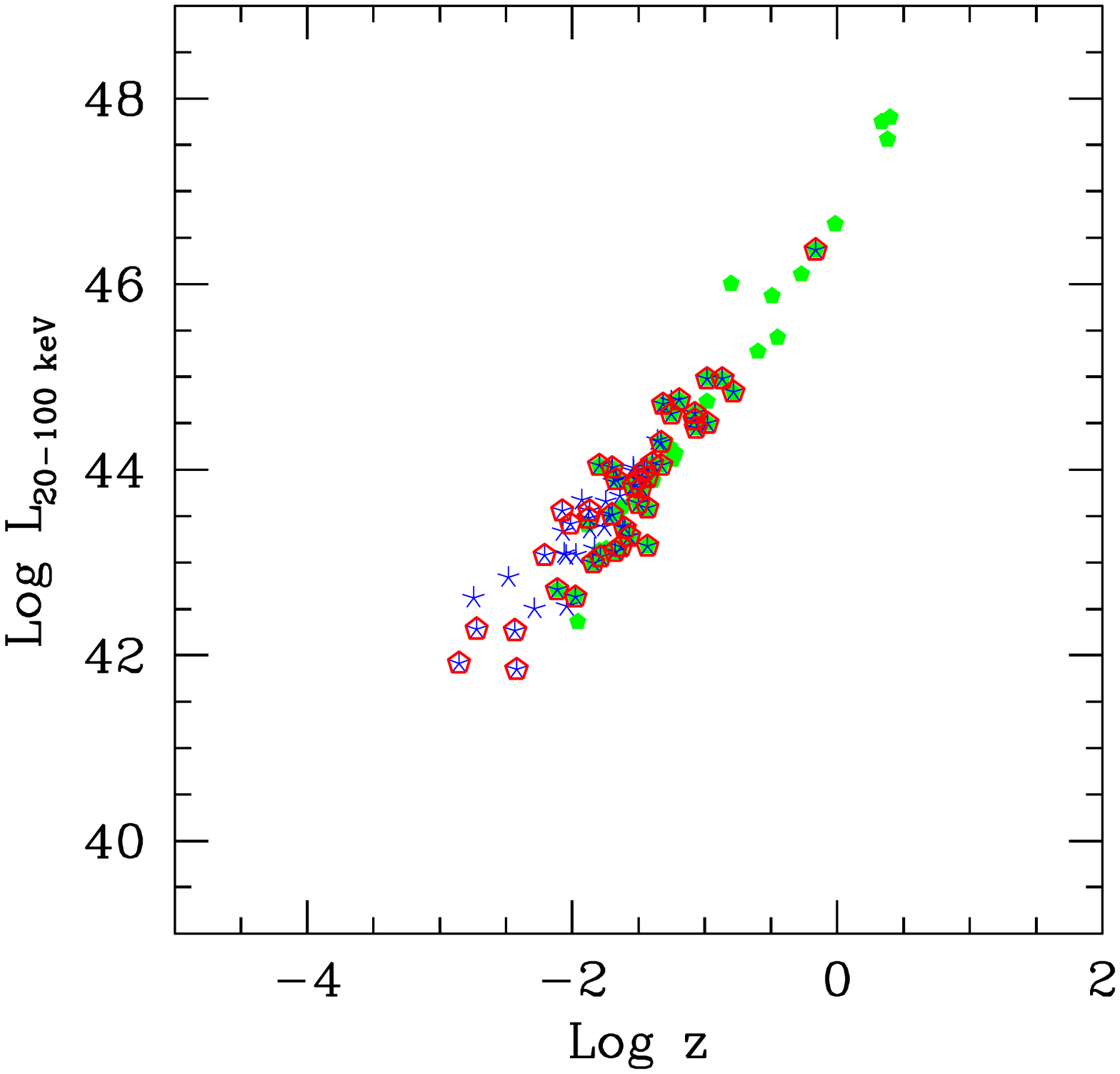}}
\vspace{-2.3cm}
\caption{Hard X-ray luminosity (20--100 keV) vs. redshift logarithm. Green dots represent  the sources that are not observed at 22\,GHz, blue stars are observed sources, and red open polygons are detected sources. Left panel: Total sample of 380 AGN. Right panel: Complete sub-sample of 87 AGN.}
\label{zlum}
\end{center}
\end{figure*}

In Fig.~\ref{zlum} we plot the hard X-ray luminosity as a function of redshift (in logarithmic scale) for the total sample (left panel) and the complete sample (right panel). We divide the AGN into those that are not observed at 22\,GHz, those that are observed, and those that are detected. This figure clearly shows that objects at high redshift and high hard X-ray luminosity are almost not covered by maser observations. Furthermore, we compared the redshift and luminosity distributions of  the total and complete samples in order to confirm whether they belong to the same parent population (null hypothesis). A Kolmogorov–Smirnov (KS) test results in p-values of 0.10 ($z$) and 0.59 (L$_{20-100 keV}$), therefore the null hypothesis cannot be rejected at the 1\% level. This suggests that the two samples might be considered statistically equivalent, that is, likely affected by similar biases. We performed the same test and only considered the observed sources in the two samples. The KS test results in p-values of 0.75 ($z$) and 0.41 (L$_{20-100 keV}$), and again, the null hypothesis cannot be rejected at the 1\% level. 
We conclude that the total and complete samples of hard X-ray selected AGN with water maser observations are representative of the local Universe and thus constitute an ideal set in which to study water maser occurrence in nearby super-massive black holes.

\section{Observations, data reduction, and results}

\subsection{GBT observations}

We observed the 6$_{16}$--5$_{23}$ transition of ortho-H$_2$O (rest frequency 22.23508\,GHz) towards 28 galaxies of the complete INTEGRAL sample with the GBT between March 2010 and January 2011 (projects AGBT10A-042 and AGBT10C-012). We used the 18--22\,GHz dual-beam receiver in nod mode and kept one of the two beams alternately on-source during integration. The GBT spectrometer was configured with two 200\,MHz IFs offset by 180\,MHz for a total coverage of 380\,MHz (corresponding to 5100\,km\,s$^{-1}$ at 22\,GHz). The first spectral window was centred at the frequency corresponding to the recessional velocity of each galaxy, and the second was offset to the red. This setup yielded a channel spacing of 24\,kHz ($\sim$0.3\,km\,s$^{-1}$ at 22\,GHz) per spectral window. We reduced and analysed the data with \textsc{gbtidl}\footnote{\url{http://gbtidl.nrao.edu}}. Flux calibration was performed using standard routines and applying the default zenith opacity and gain curve. The estimated uncertainty of the absolute flux calibration is $\sim$20\% (for details, see the guide for calibrating GBT spectral line data using \textsc{gbtidl}\footnote{ \url{https://www.gb.nrao.edu/GBT/DA/gbtidl/gbtidl\_calibration.pdf}}).

\subsection{Effelsberg observations}

On April 15 and 16, 2011, we used the Effelsberg 100 m telescope to search for 22\,GHz water maser emission in seven galaxies of the complete INTEGRAL sample (3C\,111, IC\,4329A, IGR\,J16482, 2E\,1739, IGR\,J17513, IGR\,J21247, and BL\,LAC). We employed the 1.3\,cm primary focus (PFK) receiver (17.9–26.2\,GHz) with a fast Fourier transform (FFT) spectrometer encompassing 100 MHz and 16384 channels. This setup yielded a channel spacing of 6.1\,kHz, corresponding to 0.08\,km\,s$^{-1}$ at 22.2\,GHz. 
We observed in position-switching mode, with the off-position offset by 15 arcminutes in right ascension. Signals from individual on- and off-source positions were integrated for 120\,s each. The data were reduced using the \textsc{gildas} software package (e.g. Guilloteau \& Lucas \citeyear{guillo}). To convert the measured signal from counts into antenna temperature, we used the tabulated values of the noise diode in K. We then applied the normalised gain curve and multiplied by the standard value of the sensitivity\footnote{Calibration information for the 1.3\,cm PFK receiver is reported in the Effelsberg Wiki page \url{https://eff100mwiki.mpifr-bonn.mpg.de/doku.php.}}. The uncertainty of this flux calibration was derived by applying the same procedure to continuum-pointing scans of NGC\,7027, and it is estimated to be $\sim$30\%.

\subsection{Results}

During our survey, a new water maser was detected with the GBT in the narrow line Seyfert 1 (NLSy1) galaxy IGR\,J16385-2057 on March 28, 2010. This discovery, and hence the line profile and the characteristics of the water maser emission, has been anticipated in a previous paper by our team, which was focused on water maser emission in NLSy1 galaxies \citep{tarchi11b}. Here, we report in Table 1A the isotropic line luminosity. In Table 1A we also list the 1$\sigma$ rms and the upper limit on $L{_{\rm H2O}}$ for the remaining 33 targets, with the exception of 3C\,273. The strong radio continuum emission of the blazar jet (S$_{22}$=27-43 Jy, e.g. Gear et al. \citeyear{gear94}) affects the GBT spectral baseline, which is visible as strong ripples. These prevented us from estimating a reliable rms for this source and from assessing the presence (or absence) of an emission line. As a consequence, 3C\,273 is labelled 'not observed' in Table 1A.

Although we were unable to reach the full coverage of the complete sample at 22\,GHz, we were able to increase the number of sources with water maser observations from the initial 31 with data in the literature to 65 (34 from our own survey). This increases the coverage from 36\% to 75\%.

Within the sample of 34 objects that were observed for the first time, the maser detection rate is rather low (1 of 34, or $<$6\%), but this is likely due to the optical classification of the observed sources: most objects (26) belong to the type 1 classification, and only 8 sources are of type 2. The one detected maser belongs to the class of NLSy1, which indeed seems to have a high probability of hosting maser emission \citep{tarchi11b}. In this respect, the fraction of detected maser within these newly observed AGN is consistent with the average fraction of detected masers in type 1 objects (see next Sect.).
We also searched for further biases that might be introduced by the lack of full radio observation coverage of the complete sample in addition to the known bias against high redshift and luminosity. We compared the distribution of X-ray absorption and position in the sky of the observed and unobserved sources. The test did not reveal significant differences between the two sub-samples in terms of absorption (KS p-value=0.09) or position in the sky (KS p-value=0.23).

\section{Maser fraction at high energies}

Out of 380 objects in the total sample, 193 have been observed at 22\,GHz. This is only 51$\pm$4\% of the sample. In the complete sample, 65 out of 87 objects were observed at this frequency. This provides a coverage of almost 75\%.

\begin{table*}
\begin{center}
%\label{table:table1}   
% \begin{minipage}{140mm}
\centerline{Table 1: Summary of the detection fractions for different samples and sub-samples.}
 \begin{tabular}{l l l}
\hline\hline
Sample (Number of AGN observed at 22GHz)            &  DF (Detection fraction in \%) &  DF (Detection fraction in \%) range \\
\hline
INTEGRAL total (285)    & 15.0$\pm$2.8                      &  7.6-57.0\\
INTEGRAL complete (65) & 18.5$\pm$5.3                    & 13.8-39.0  \\   
INTEGRAL Sey1.8-2 total (103)& 22.0$\pm$4.7                      & 13.5-51.0 \\
INTEGRAL Sey1.8-2 complete (29) & 31$\pm$10              & 27-41 \\
INTEGRAL Sey1-1.5 total (74)  & 2.7$\pm$1.9                     & 1.4-51.0\\
INTEGRAL NLSy1 total (9)     & 33$\pm$19                       & 20-60 \\
Swift/BAT 70M (285)      & 12.6$\pm$2.1                    &  -    \\
Swift/BAT 9M (114)      & 14.9$\pm$3.6                    & 11-37  \\
Swift/BAT 9M Sey2 (51) & 25.5$\pm$7.0                      & 20-41  \\ 
DS optical     (89)    & 23.6$\pm$5.1                    & -    \\
DS optical Sey2(71)     & 26.8$\pm$6.1                     & -  \\
 
\hline
\hline

\end{tabular}                                           
                
\end{center}                                                                                                     
\small                                                                                                                   
\end{table*}
Considering the total set of INTEGRAL AGN reported in Table 1A, we found that out of 193 galaxies observed at 22\,GHz, 29 have been reported as maser sources; this represents 
a detection rate of 15$\pm$3\%. Errors on the fractions were calculated as $\sqrt{N_M}/N_O$, where $N_{M}$ is the number of detected maser sources and $N_{O}$ is the number of observed sources. To take unobserved sources into account and thus  provide a range of values for the entire sample, we can take two extreme approaches and assume that if pointed, all not yet observed INTEGRAL AGN are found to be undetected at 22\,GHz (lower range), or alternatively, that all are detected (upper range). Under these conditions, we find that the detection rate for the whole sample ranges from 8 to 57\%. This a rather wide range that nevertheless tells us that the detection rate in the worst case is higher than is generally obtained using large samples of optically selected galaxies (Sect. 1). 
Out of 29 objects with maser detection, the 22\,GHz luminosity is below 10 L$_{\odot}$ for 6 objects (i.e. Mkn\,3, NGC\,4051, NGC\,4151, Mkn\,766, Cen\,A, and NGC\,6300) and they therefore qualify to be kilo-maser objects. However, some of these sources were imaged at high resolution and their maser emission was found to be located within the nuclear region of the host galaxies. Therefore, given that all our sources are hard X--ray emitters and can confidently be associated with an accreting supermassive black hole, we confirm that kilo-maser emission may not only be associated with star formation, but also with AGN activity, as has been suggested by Tarchi et al. \citeyear{tarchi11a}.

When we instead consider the complete sample of INTEGRAL sources that is highlighted in Table 1A, 65 AGN are observed at 22\,GHz, of which 12 are detected and only 22 are unobserved.
In this case, the detection rate is 19$\pm$5\% and the range of possible values lies between 14 and 39\%.

Because water maser emitters show a preference to be found in type 2 AGN, we also restricted our analysis to Seyfert 2 galaxies alone (including intermediate types 1.8-1.9), which are present either in the total or in the complete INTEGRAL samples. In this case, we find that the sample coverage was 68\% (103 AGN observed, 23 detected, and 60 unobserved) and 88\% (29 AGN observed, 9 detected, and 5 unobserved) for the total and complete sample, respectively: we therefore estimate a detection rate of 22$\pm$5\% (range values from 14 to 51\%) and 31$\pm$10\% (range values from 27 to 41\%) for each of these two samples. 

These fractions are even more remarkable when they are compared to those of Seyfert 1 (in this case, also including intermediate types 1.2-1.5): 74 objects have water maser observations of the type 1 AGN, but only two (NGC\,4151 and NGC\,3783) have been detected. This implies a detection fraction of about 3\%, which is similar to what we found within the MCP (see also K{\"o}nig et al. \citeyear{konig}). 
Interestingly, the detection rate of NLSy1, which are also broad-line AGN but have peculiar characteristics at multi-frequencies with respect to standard broad line AGN (see e.g. Panessa et al. \citeyear{panessa}), is comparable to those of Seyfert 2. Our sample includes 9 NLSy1 with 22\,GHz measurements and 3 water maser detections (NGC\,4051, Mkn\,766, and IGR\,J16385-2057), which implies rates close to 30\%. This confirms previous results obtained by Tarchi et al. (\citeyear{tarchi11b}) in a dedicated study of this type of AGN, in which the authors suggested an outflow origin for water maser emission. \citet{mathur} proposed that NLSy1 sources might be young AGN residing in rejuvenated galaxies; alternatively, their peculiarities might be explained in terms of an orientation effect that is caused by the pole-on observation of their broad-line region \citep{decarli}. The link of these two interpretations to the maser phenomenon in these peculiar objects is still unclear, but confirmation of high detection rates in NLSy1 indicates that this question requires more in-depth studies.

To consolidate our overall results, we also compared the above rates with those obtained from the 70-month \citep{baum} and 9-month \citep{tueller} Swift/BAT samples. These two samples were selected to be almost comparable in size with the INTEGRAL total and complete sub-samples.

The 70-month BAT survey provides the list of all objects detected by the instrument during the first 6.8 years of the Swift mission and covers 90\% of the sky at a sensitivity level of 1.3 $\times$ 10$^{-11}$ ergs s$^{-1}$cm$^{-2}$ in the 14--195 keV band.
The sample contains a large fraction of unclassified sources that may turn out to be AGN after proper follow-up work; thus our search for maser detection provides only an indication of the level of maser occurrence  in this large BAT sample. To search for water maser emission in this set of hard X-ray selected AGN, we were helped by three high-school students during a stage performed at OAS/INAF in Bologna in 2016. As for the INTEGRAL sample, we searched the available archives, such as the MCP and the literature, for reports of water maser observations and detections for all 822 AGN reported in the BAT survey. Altogether, we found that only 285 objects of this sample (only 35\%) have been observed at 22\,GHz and 36 objects have been detected.
The detection fraction is therefore about 13$\pm$2\%, which perfectly agrees with our INTEGRAL results. 
Most of the detections overlap the INTEGRAL detections; the additional sources are NGC\,235A, UGC\,3157, VII Zw\,073, NGC\,3393, CGCG\,164-019, MKN\,78, UGC\,5101, and  M\,82 considering that IGR\,J16385-2057 and NGC\,6926 were only detected by INTEGRAL. We note that in the case of M\,82, the emission above 10 keV is dominated by a few ultra-luminous X-ray sources (ULXs) with a minor contribution from lower luminosity X-ray binaries \citep{vulic} and can therefore not be attributed to AGN activity.

The Swift/BAT 9-month catalogue contains only 154 sources (all of which are identified and optically classified as AGN) and covers 74\% of the sky (only the sky above $\pm$15 degrees in latitude was considered to avoid contamination by galactic objects) at a flux threshold of 5$\times$ 10$^{-11}$ ergs s$^{-1}$cm$^{-2}$ in the 14--195 keV band. In this catalogue, 114 objects have been observed at 22\,GHz and 17 are detected (listed in Table 2A). Only 40 AGN have no observational coverage at the waveband of interest here. The detection rate is 15$\pm$4\%. Applying the same exercise as for the INTEGRAL samples (i.e. assuming all unobserved sources to be either detected or undetected at 22\,GHz), we estimate a possible range of values between 11 and 37\%. When we restrict this to type 2 AGN alone, 51 of 65 objects in this sample have been observed, and13 objects display maser emission. This provides a detection rate close to 26$\pm$7\%. In this case, the possible range of values is estimated to vary from 20 to 41\%, which again fully agrees with the estimates obtained from the INTEGRAL samples.

\begin{figure*}
\begin{center}
\parbox{18cm}{
\includegraphics[width=0.3\textwidth,height=0.3\textheight,angle=0]{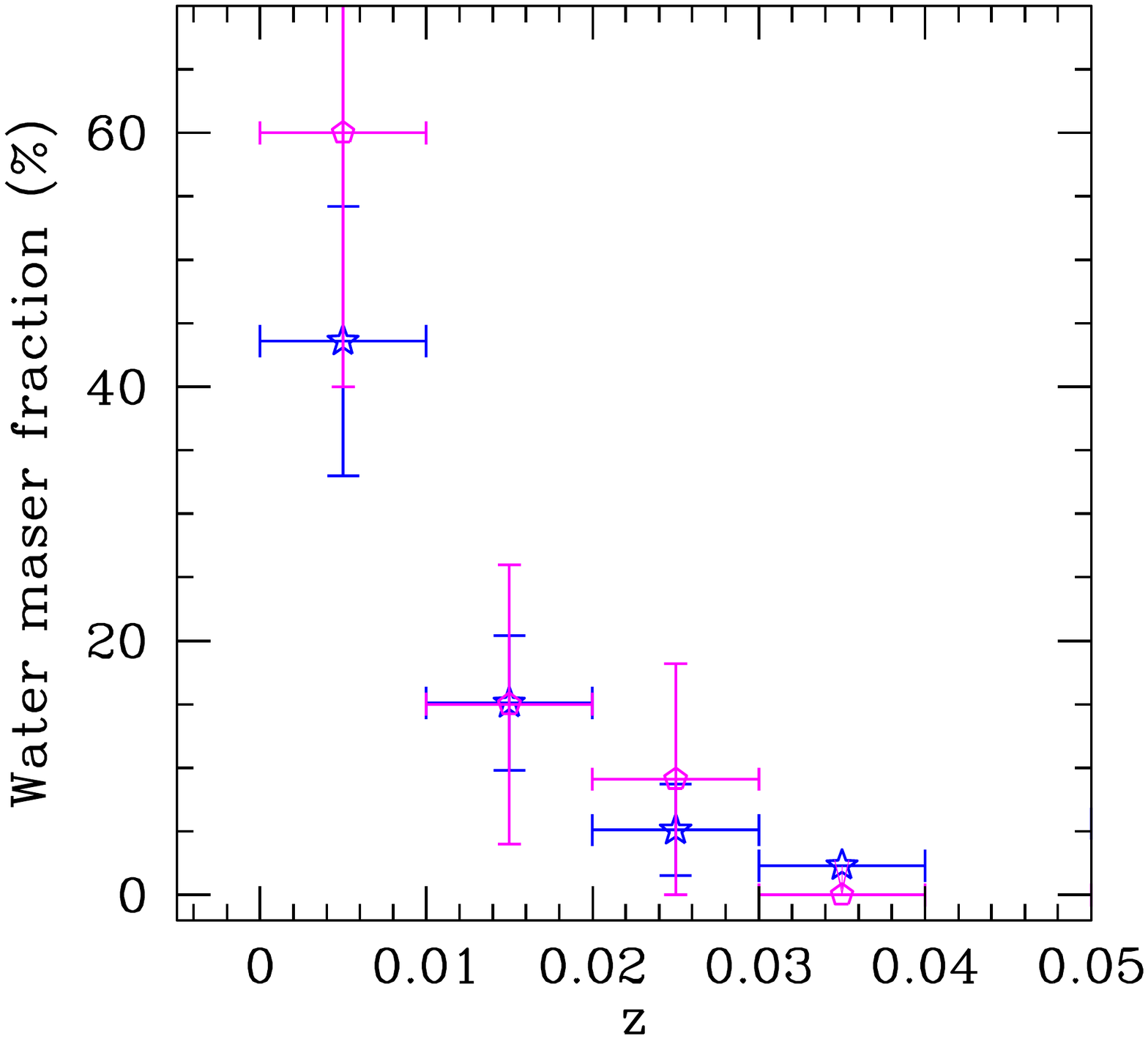}
\includegraphics[width=0.3\textwidth,height=0.3\textheight,angle=0]{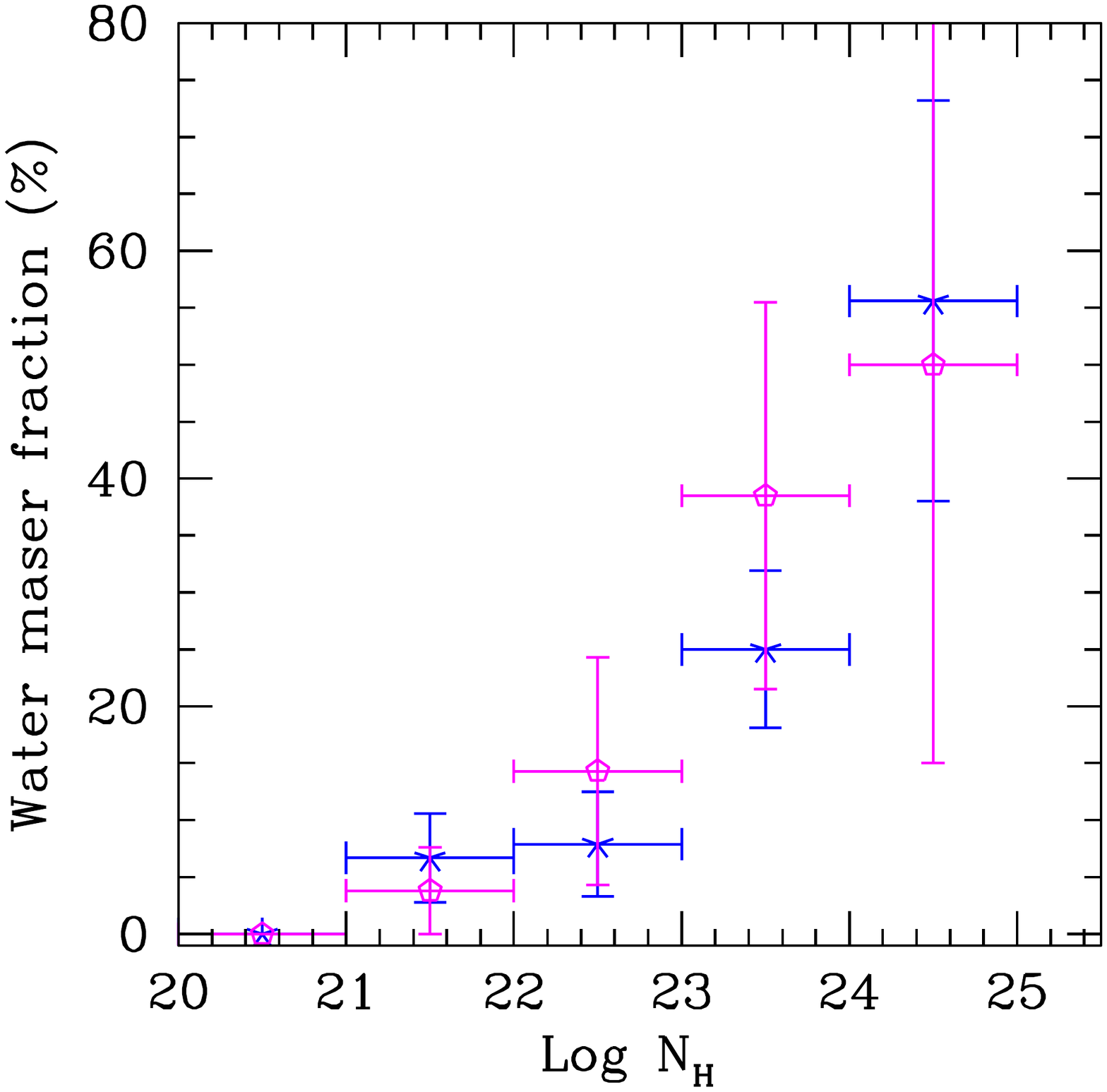}
\includegraphics[width=0.3\textwidth,height=0.3\textheight,angle=0]{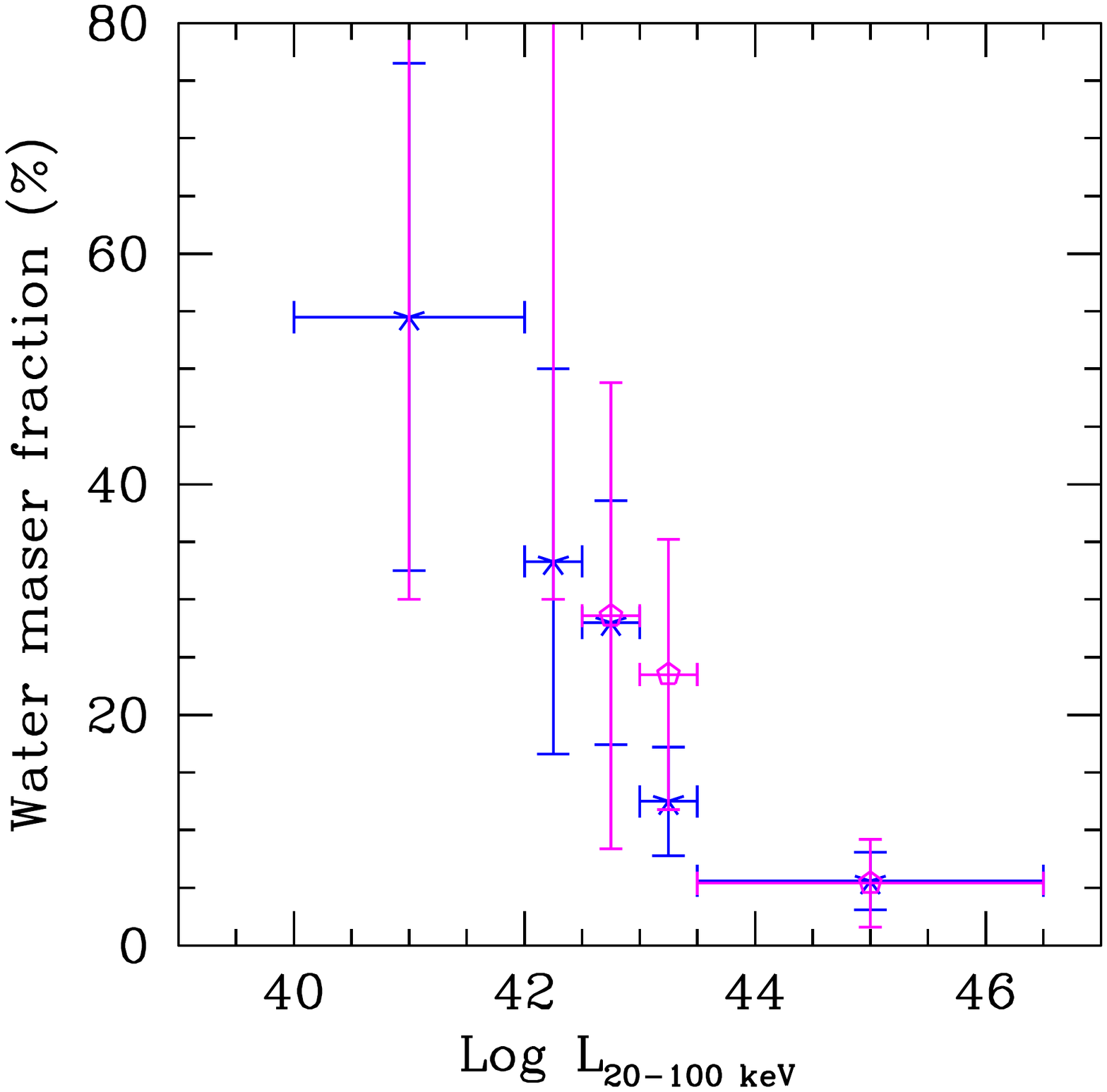}
}
\vspace{-1.5cm}
\caption{Fraction of detected water maser emission vs. redshift limited to z$<$0.04, X-ray column density in cm$^{-2}$ (central panel), and 20--100 keV luminosity in ergs s$^{-1}$ cm$^{-2}$ (right panel). Blue starred points represent the total sample, and magenta polygons show the complete sub-sample.}
\label{fraction}
\end{center}
\end{figure*}

Finally, we compared our results, in particular, those obtained for type 2 Seyferts, with the optical data set of AGN discussed by \citet{diamond}. These authors have compiled a complete sample of 89 Seyfert galaxies, made of 18 type 1 (1-1.5) and 71 type 2 (1.8-2) AGN, all within a distance of 200 Mpc. The entire sample has been covered by water maser observations and is therefore a reference catalogue for this type of studies. There are 21 masers (listed in Table 3A) and 68 non-maser sources in this sample, providing a detection rate of 24$\pm$5\%. When we restrict the estimate to Seyfert 2 galaxies alone, the detection rate increases only slightly to 27$\%$, again in full agreement with the estimate provided in this work. The highest detection rate found in this sample appears to contrast with estimates obtained using other optically selected samples of AGN (e.g. \citet{zhu} quote a detection rate of almost 8\%\ for Seyfert 2 galaxies), but this may be due to the fact that the \citet{diamond} sample is biased in favour of close-by AGN, mostly of type 2, similarly to the sample that was investigated in Panessa \& Giroletti (\citeyear{pg}).

The detection fractions for all samples considered in this work are summarised in Table 1, which clearly shows that hard X-ray catalogues provide a significant boost of the maser detection fraction with respect to large optical surveys (e.g. Zhu et al.  \citeyear{zhu}; Braatz et al.  \citeyear{braatz18}): they reach values of at least 15-25$\%$.
These values are also higher than those obtained by specifically tuning the AGN selection in the IR band using different criteria and combining them, as recently proposed by \citet{kuo18}. Furthermore, the hard X-ray selection, in addition to providing a catalogue of galaxies with a high probability of maser detection, also provides a set of sources with clear evidence of AGN activity and therefore negligible or null contamination from star-forming objects.

\section{Improving the detection probability}

In Fig.~\ref{fraction} we plot the water maser detection rate in our total sample and in the complete sub-sample as a function of redshift (left panel), X-ray nuclear absorption (middle panel), and 20--100 keV hard X-ray luminosity (right panel). For the distribution of the water maser fraction as a function of redshift and column density, objects were grouped so that about the same number of observed sources is shown in each bin. For the distribution in luminosity, this was more difficult to achieve while still maintaining a reasonable number of bins, but the uncertainty related to this choice is reflected in the error associated with each bin.

It is evident that water maser detection decreases as a function of redshift from about 40-60\% at low redshifts to a few percent at higher distances (above z=0.015), in agreement with the distance bias discussed in Sects. 2 and 3. In contrast, the detection fraction increases for higher X-ray column densities from a few percent to 25-40\% at a threshold of 10$^{23}$ cm$^{-2}$ and reaches 56$\pm$18\% above the Compton-thick regime in the case of the total sample (50$\pm$35 for the complete sample). This again confirms that the water maser detection is favoured among heavily absorbed AGN.
The hard X-ray luminosity also appears to play a role, although in these cases, the error bars are larger and the scarce observations of highly luminous sources may affect this result: a decline in detection fraction is seen from low to high luminosities, which may be an indication that an extremely luminous nuclear environment does not favour maser emission.
As discussed in Castangia et al. (\citeyear{cast13} and references therein), for high nuclear bolometric luminosities or environments that are particularly heavily exposed to strong X-ray radiation, H$_{2}$O maser emission at sub-parsec distances from the nucleus of the galaxy may be hampered because the ISM is mostly atomic and/or the dust grains (where water is thought to be often formed) are destroyed. While maser emission can still be produced at larger distances, the innermost (hundredths of parsecs) masing action would be prevented if this scenario is correct. In addition, an increased bolometric luminosity might decrease the maser emissivity by reducing the difference between gas and dust temperatures \citep{kuo18}, on which the volume rate of maser photon production significantly depends \citep{gray}.

The fractions for the total and the complete sub-samples are consistent within the errors and the trends are confirmed for the two samples. However, the statistics is limited by the small number of data sets, especially for the complete sub-sample.

\section{Maser type of INTEGRAL AGN}

Most maser detections reported in this work have been discussed in the literature and their maser type has been analysed in previous works (see type and relative references in the notes of Table 1A). In the following and in Table 1A, we  considered as disc, outflow, or jet maser sources for which the  maser class is either fully assessed or just suggested on the basis of observational results; for these last objects, only follow-up, mainly interferometric VLBI, continuum, and spectral line studies can confirm a water maser association with AGN activity and maser type.

Only 6 sources in the sample of 29 maser detections (IGR\,J05081+1722, NGC\,3081, NGC\,3783, NGC\,5643, NGC\,6300, and ESO\,103-G35) have no associated maser classification; in Appendix B we attempt to provide some indication of the possible nature of these sources and implicitly discuss their most likely maser type.  Table 1A shows that the water maser classification generally refers to one, or in some cases, two components, such as a disc plus outflow or jet.

When we exclude these 6 sources from our sample of AGN with 22\,GHz detection, we note that similar numbers (12-12) of objects have a disc (or evidence of disc) and a jet (or evidence of jet) water emission; outflow or evidence of outflow emission is present in 7 sources. Despite the uncertainties involved in water maser classification, it is evident from the present sample that all types of masers are likely associated with INTEGRAL AGN and that discs masers are not necessarily the dominant type. Thus hard X-ray surveys also offer the opportunity of probing masers of different types.

Finally, we note that the maser type of all optical narrow-line AGN (including type 1 and NLSy1) is likely related to jet and/or outflow emission, and none is apparently associated with accretion discs. Type 2 AGN instead seem to display all types of water masers: out of 24 Seyfert 2 that are detected, the emission in 12 is partly or entirely associated with a disc, in 8 with a jet, and only in 3 with outflow. Maser sources associated with edge-on ($i$ = 90$\pm$10 degrees, see Sect. 8) discs are most likely, and not surprisingly according to the unified model, found in type 2 AGN.

\begin{figure*}
\begin{center}
\parbox{18cm}{
\includegraphics[width=0.3\textwidth,height=0.3\textheight,angle=0]{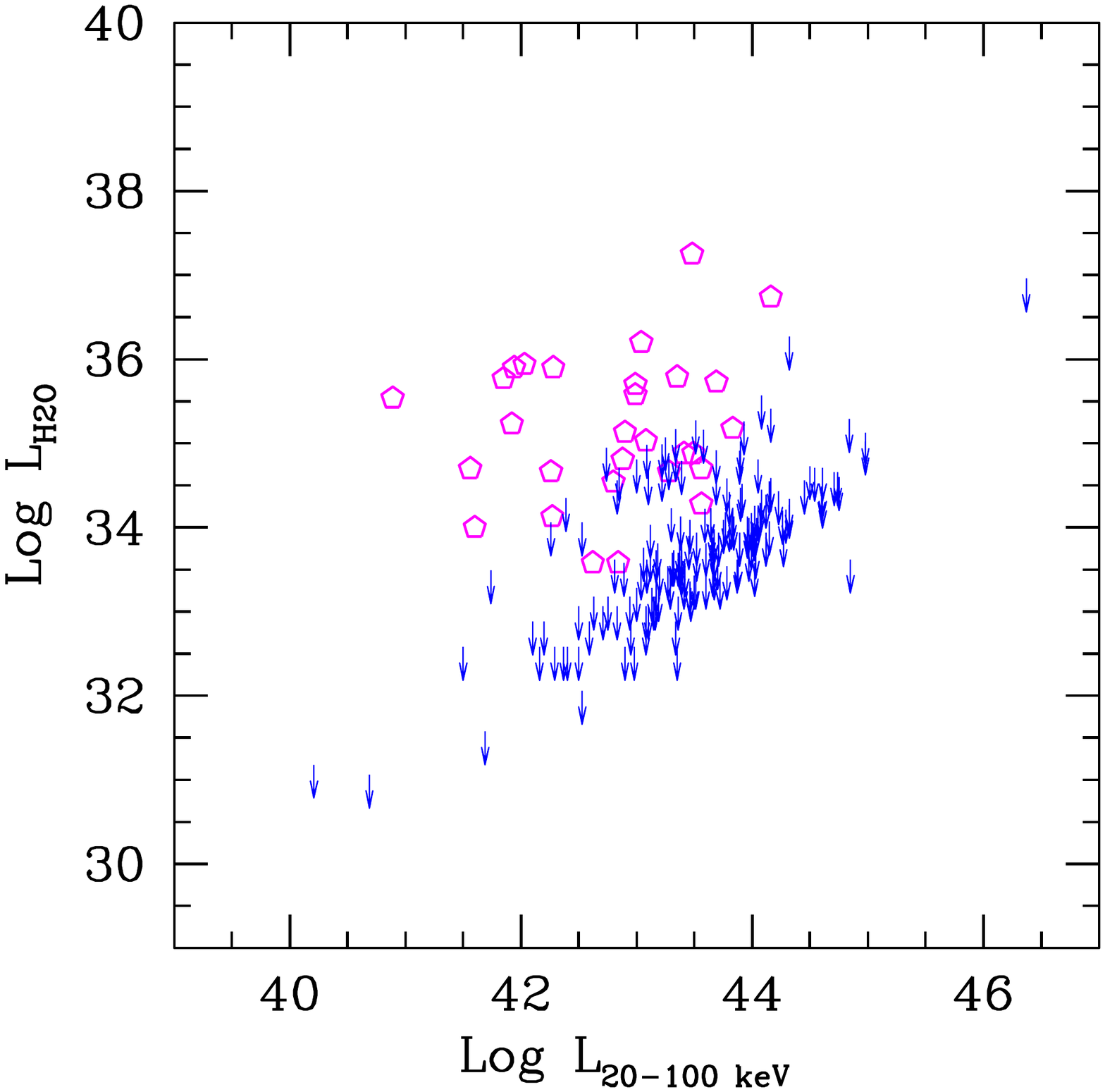}
\includegraphics[width=0.3\textwidth,height=0.3\textheight,angle=0]{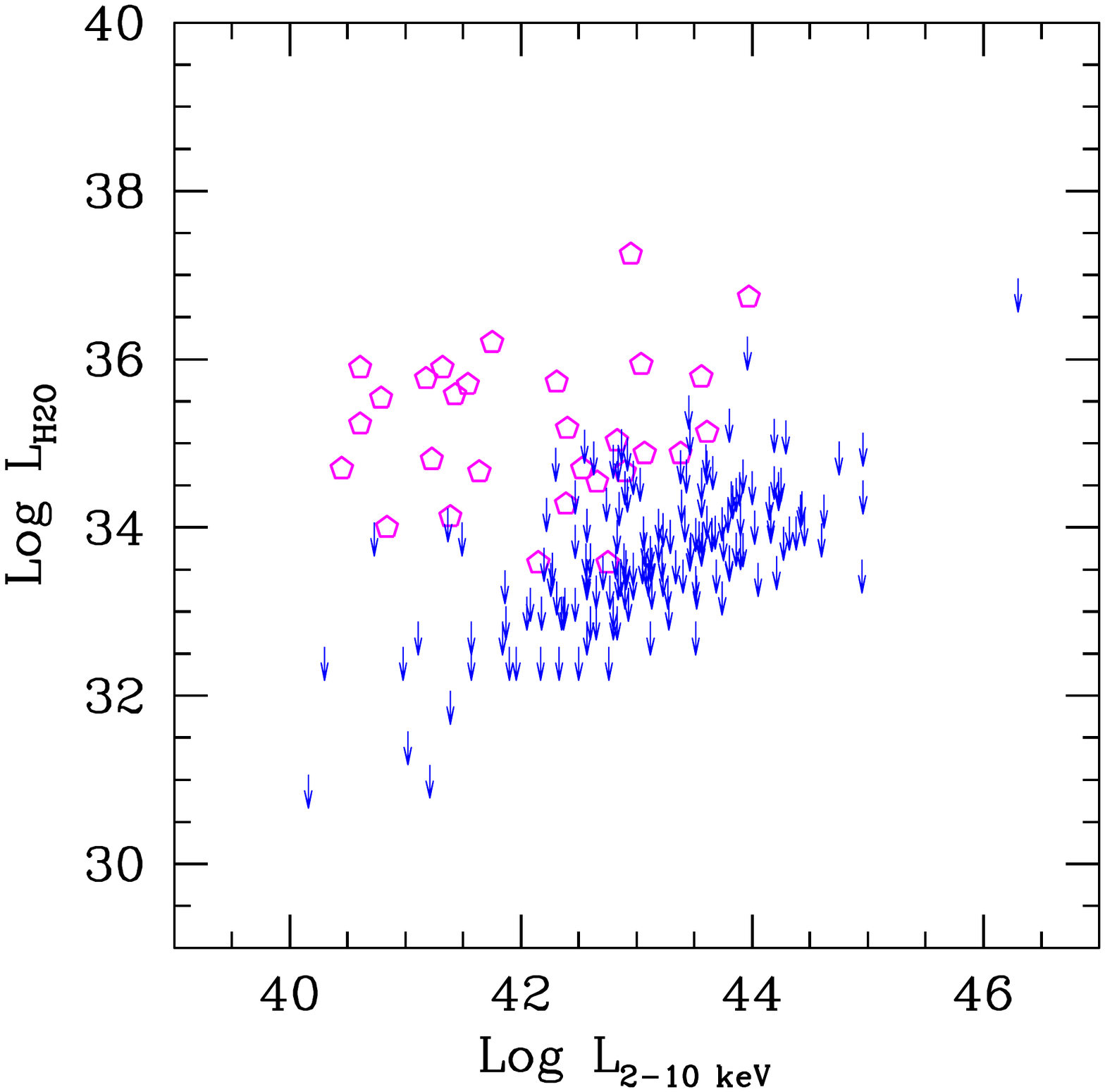}
\includegraphics[width=0.3\textwidth,height=0.3\textheight,angle=0]{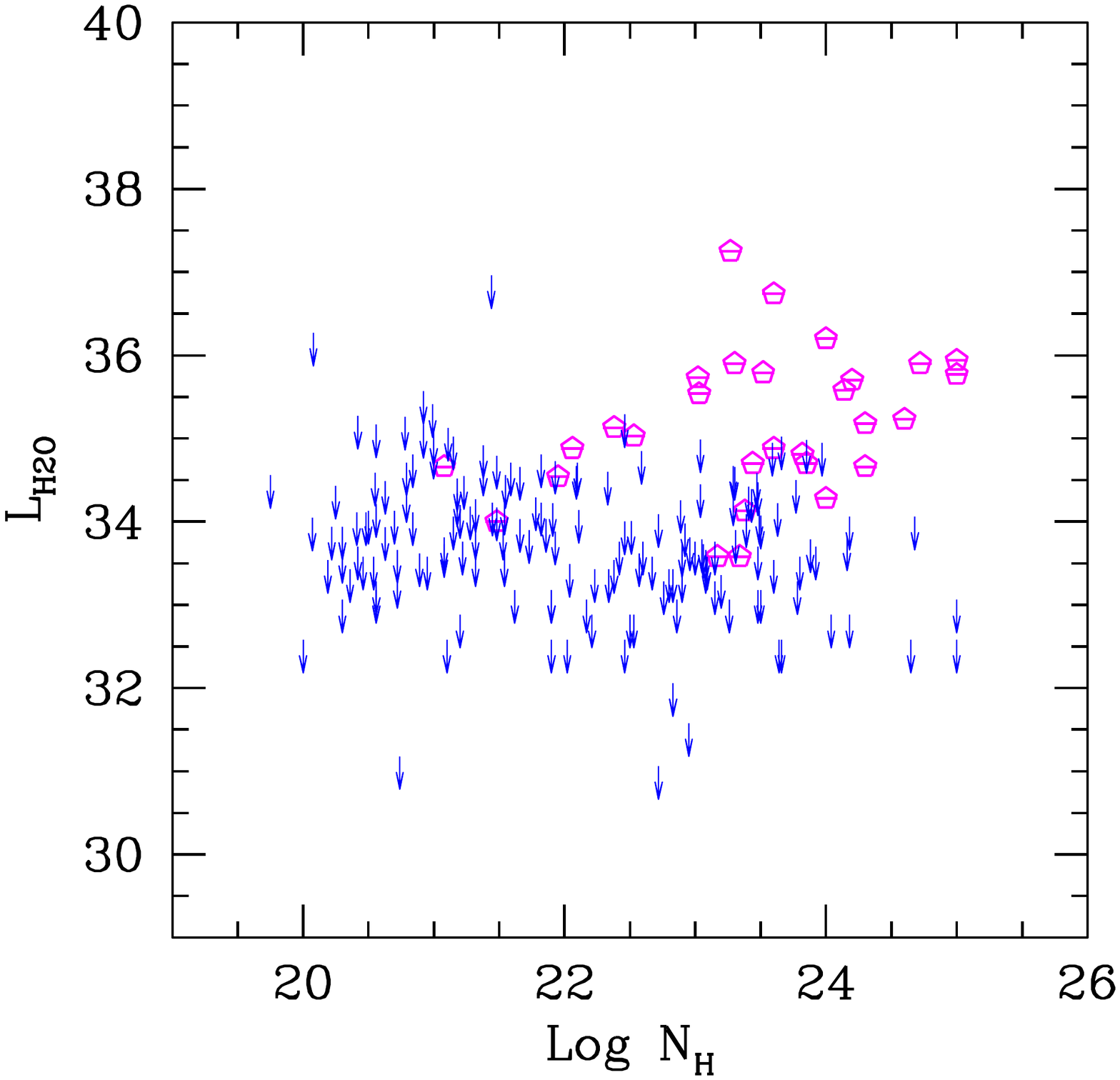}
}
\vspace{-1.5cm}
\caption{Logarithmic water maser luminosity vs. 20--100 keV (left panel) and 2--10 keV logarithmic luminosities (central panel), expressed in ergs s$^{-1}$ cm$^{-2}$. Water maser luminosity vs. the logarithmic X-ray column density in cm$^{-2}$ (right panel). Empty magenta polygons are water maser detected sources, and blue arrows represent the upper limits to the water maser luminosity.}
\label{corr}
\end{center}
\end{figure*}

\section{Maser versus non-maser INTEGRAL AGN}

One main question that is still unanswered in extragalactic maser astrophysics is related to the conditions that lead to maser emission in only a fraction of AGN. iIt is therefore reasonable to ask whether water maser galaxies have special intrinsic properties in terms of X/hard X-ray luminosities and absorption compared to apparently similar galaxies without detected maser emission. 

According to theory, high-energy radiation from the central part of an AGN could heat the circumnuclear gas temperature to values suitable for maser emission \citep{neufeld}; in this case, we expect a relationship between the maser luminosity and the X-ray/hard X-ray luminosities. \citet{kondratko06b} studied a sample of 30 water maser AGN and indeed found such a relation (L$_{X}$ $\propto$ L$_{H2O}^{0.5}$), where L$_{X}$ is the unabsorbed X-ray luminosity in the 2--10 keV band; the significance of this correlation improved when the sample was limited to disc masers. However, the relation still presented a large scatter, likely due to a dependence on different parameters, such as the mass accretion rate, the ratio of X-ray to bolometric luminosity, and the well-known X-ray and maser variability (in the latter, this typically is about some dozen percent, e.g. \cite{maloney} and references therein). This relation seems to be weak or absent in more recent studies even when only disc maser sources were considered, which are the sources in which the correlation is expected to be stronger \citep{cast13}.

A more direct estimate of the AGN radiation field is provided by the hard X-ray luminosity, which is the least affected in terms of nuclear absorption; this information is available for all our objects and has never been employed before in a correlation with the maser luminosity. 
In Fig.~\ref{corr} (left and middle panels) we therefore plot the isotropic water maser luminosity as a function of the 20--100 keV and 2--10 keV observed luminosities. When the regression analysis is only applied to the detected sources, the Spearman Rho correlation coefficients are 0.16 and 0.10, providing a two-tailed probability of 0.42 and 0.59, respectively. Therefore the association between the two variables should not be considered statistically significant. The sources of scatter as discussed in \citet{kondratko06b} analogously apply to our relations. 

Finally, we note here also that the range of hard X-ray and X-ray luminosities of detected masers cover a similar interval as non maser AGN, therefore no evident luminosity threshold could be identified as maser activator above L$_{2-10 keV}$ $\sim$ 10$^{40}$ ergs s$^{-1}$. In addition, the sensitivity of 22~GHz surveys is no limit for maser detection because luminosity upper limits are also found at a factor of $\text{about}$ 10 to 100 below detections.

As discussed by \citet{zhang06}, a correlation between water maser luminosity and X-ray absorption (approximately L$_{H2O}$ $\propto$ N$_H^{3}$) is also expected for idealised saturated maser emission (assuming no velocity gradients in the maser region). In this case, the value of the exponent is determined by the luminosity that linearly increases with the column density and the surface of the masing cone, which grows with the square of its lengths (e.g. Kylafis \& Norman \citeyear{kylafis}). In Fig.~\ref{corr} (right panel), we display the isotropic water maser luminosity as a function of the X-ray absorption for maser and non-maser sources. When we again consider only detected sources, the resulting Spearman Rho correlation coefficient is 0.37 with a derived two-tailed probability of 0.05, suggesting that the association between the two variables might only be marginally considered statistically significant. This again confirms that the X-ray obscuring medium is associated with the masing material.

While an interpretation based on the different maser types would be interesting, it would reduce the number of sources that are tested because variety of maser types in our sources is great. This would weaken the significance of these relations. In addition, the high complexity of the different maser components complicates the interpretation of these correlations.

\section{Water masers in Compton-thick AGN}

Of the 21 objects of the total sample in the CT regime, 8 have no maser detection, 10 have a maser detection, and 3 have not yet been observed. Therefore the probability is about 56$\pm$18\% to detect maser emission in the Compton-thick AGN set selected in the hard X-ray band (see also Sect. 5). This is in line with the noticeably large fraction of water masers (50\%) found by \citet{cast19}, who studied a sample of heavily absorbed AGN, including CT sources, that were selected through a combination of mid-IR and X-ray data. We note that all CT AGN in the INTEGRAL sample show evidence of an association with discs (sometimes accompanied by jet and outflow components), except for Mkn\,3, which is tentatively associated with a jet origin only (a core plus jet component is seen in the radio continuum maps of this source, Chiaraluce et al. \citeyear{chiara}). When we exclude Mkn 3, the fraction of disc masers in CT is 50$\pm$17\%. The interesting question here is why some sources are able to develop strong maser emission while others are not. In other words, if all Compton-thick AGN are potentially water maser emitters, why are only half of them able to reach luminosities high enough (above 10$^{33}$ erg cm$^{-2}$ s$^{-1}$, see Fig. 3) for the current generation of radio receivers?

\citet{masini18} have estimated that the expected disc water maser detection fraction among type 2 Seyferts in a volume-limited survey is about 10\% to 20\%. This value has been obtained by comparing the torus and maser disc covering factors (see their equation 2.1) and assuming that a maser disc is detected when the line of sight angle ranges between 90$\pm$10$^{\circ}$ with respect to the polar axis and defining the probability of detecting a maser disc in a type 2 AGN as the ratio of the maser disc covering factor with respect to the torus one. This assumption might be inverted considering a covering factor for the CT part of the torus as derived from X-ray arguments \citep{ricci}  to be $\sim$ 23\% (which also agrees with IR arguments, e.g. H\"onig \citeyear{honig}) and using the information on the ratio between the water maser and CT covering factors to be 50\%, to finally derive the expected maser disc inclination angle to range between 82 and 87 $^{\circ}$. These values agree with the observed disc angles in well-known disc masers \citep{kuo11, konig}, which confirms the idea that the masing disc is only a portion of the total CT medium (for a sketch of the possible geometry discussed here, see Fig. 2 in Masini et al. \citeyear{masini16}). The maser covering factor could be considered as a lower limit when we assume the presence of warped discs, which is expected to increase the probability of intercepting maser emission \citep{darling}. Similarly, the known water maser variability might contribute to a possible non-detection if the masers flux decreases below the instrument sensitivity and therefore to an underestimate of the covering factor. Other effects contribute to our uncertainties in these estimates, for instance, X-ray scattering in clumpy media might dilute the true line-of-sight column density and thus prevent us from deriving unbiased orientation information \citep{ramolla}. Despite the uncertainties involved, it is important to stress that the 50\% detection fraction in Compton-thick AGN can in principle be explained in terms of a geometrical effect (i.e. proper opening angles of torus and maser disc and their relative alignment) and may not be due to peculiarities of individual objects.

\section{Conclusions}

Notwithstanding the valuable science that can be derived for AGN and cosmological studies, water megamasers are rarely found in galaxy surveys (e.g. Braatz et al. \citeyear{braatz18}). We here selected a sample of hard X--ray AGN detected above 20 keV by INTEGRAL/IBIS and searched for water megamaser emission among them in the literature and through our new dedicated observations (where one new maser detection was obtained). Of the 380 sources of the sample, only 51\% have been observed at 22\,GHz. In 15$\pm$3\% of them, a detection was made. We also considered a sub-sample of 87 sources, limited in volume and statistically complete, and found that the detection fraction increases to 19$\pm$5\%. Most observed sources are at low redshift, and this is reflected by the observed detection fraction, which decreases with increasing redshift. This likely introduces a bias in our sample.

So far, the detection rates observed in large surveys of optically selected galaxies were about a few percent, which could only be improved by carefully selecting smaller samples on the basis of IR (8-22\%; Hagiwara et al. \citeyear{hagiwara02,hagiwara03}; Henkel et al. \citeyear{henkel05}) or a combination of mid-IR and X-ray data (50\%; Castangia et al. \citeyear{cast19}). Therefore, the hard X--ray selection provides one of the highest rates ever observed so far. 

These fractions increase in type 2 Seyfert galaxies (22$\pm$5\%), in particular in CT AGN, $\text{about}$ 50\% of which host water maser discs. This clearly indicates that the X-ray obscuring gas is related to the maser dusty medium. A comparison between the covering factor of the CT obscuring medium and the fraction of water masers in CT sources confirms the idea that the masing disc might only be a portion of the CT obscuring medium and that an edge-on line of sight ($i>$ 82 degrees) is required for the water maser emission to be detected. 

A possible decrease in detection fraction is observed as the hard X-ray luminosity increases, suggesting that a highly luminous nuclear environment might not favour maser emission. However, this result can be confirmed by completing the sample observations at higher luminosity. On the other hand, no significant correlation between the water maser and X-ray and hard X-ray luminosities has been found, while the marginally significant correlation between the water maser luminosity and the X-ray column density simply reflects the connection between the X-ray obscuring and the masing media.

All types of water masers are found by the soft gamma-ray selection of sources. 
Interestingly, all of the few water masers detected in type 1 AGN are jet or outflow candidates, while in type 2 AGN all types of masers are detected. This suggests that the dusty water maser medium is not solely associated with a classical obscuring torus, but might also reside in polar outflows or jets. This implies a more complex geometry, as envisaged by recent IR interferometric studies (see H\"onig et al. \citeyear{honig}).

Overall, we conclude that hard X-ray samples of AGN provide the opportunity of significantly increasing the maser detection efficiency compared to previous surveys. Extremely high detection fractions (up to 50$\%$) can be reached by targeting type 2 or heavily absorbed AGN that are nearby and in an optimised luminosity range. The discovery of new heavily absorbed sources with the increased sensitivity of the ongoing INTEGRAL/IBIS and Swift/BAT surveys together with the wealth of new sources that the eROSITA survey \citep{merloni} will discover below 10 keV will offer the possibility to largely increase the samples for future water maser searches, and hopefully, detections.\\

{\bf Acknowledgements}\\

The authors wish to thank the referee for her/his valuable comments that considerably improved our manuscript. AT and PC would like to thank Jim Braatz for providing information on some of the maser sources, prior to publication. FP, AM, LB, AB and PU acknowledge financial support from ASI under contract INTEGRAL ASI/INAF n.2019-35-HH. We acknowledge the help of 3 high school students (Alice Bizzarri, Andrea Cocozza and Fulvio Talarico) in the analysis of the BAT 70 m sample; they all participated in a summer stage at OAS/INAF Bologna during 2016. This project, being partly conducted by amateur astronomers under the supervision of professional scientists, represents a nice example on how citizen science work can help in dealing with a large data set.


\appendix

 %\newpage
%\section{Appendix}
\section{Tables with the total sample and detection fractions in the Swift/BAT and Diamond-Stanic samples}

Table 1A lists all 380 INTEGRAL/IBIS AGN with their optical coordinates, redshift, class, hard X-ray (20--100 keV) flux, X-ray (2--10 keV) flux, X-ray column density, a note to indicate whether the source was observed at 22\,GHz, which maser type was detected, and respective references. Finally, for sources for which maser emission was detected, we also list the reported water maser isotropic luminosity and the reference to the maser data; for objects observed at 22
GHz but without a detected maser, we report the 1$\sigma$ rms and an upper limit to the maser luminosity.
Objects belonging to the complete sample are highlighted in boldface in Table 1A for clarity. 

In Table 2A we report the 17 detected sources from the Swift/BAT 9 catalogue with their names, optical classification as type 1 or 2 AGN, and coordinates. Analogously, in Table 3A, the 21 AGN detected at 22\,GHz from the \citet{diamond} sample are reported.

\begin{table*}
\begin{center}
%\twocolumn
% \begin{minipage}{140mm}
\centerline{Table 2A: AGN in the BAT 9-month survey sample with water maser detections}
\begin{tabular}{l r r r  }
\hline
\hline
Name (class)& RA(J2000), Dec(J2000) & Name (class) & RA(J2000), Dec(J2000) \\
\hline
NGC235A(2) &00 42 52.81,$-$23 32 27.7 & NGC4945(2)&13 05 27.28,$-$ 49 28 04.4\\
Mkn348(2)  &00 48 47.10,+31 57 25.0   &NGC5128(2) &13 25 27.61,$-$43 01 08.8 \\
MKN3(2)    &06 15 36.31,+71 02 14.9   &NGC5506(2) &14 13 14.87,$-$03 12 27.0 \\
NGC3081(2) &09 59 29.54,$-$22 49 34.6 &NGC5728(2) &14 42 23.90,$-$17 15 11.0\\
NGC3783(1) &11 39 01.78,$-$37 44 01.7 &NGC6240(2) &16 52 58.97,+02 24 01.7\\
NGC4051(1) &12 03 09.62,+44 31 52.8   &NGC6300(2) &17 16 59.47,$-$62 49 14.0\\
NGC4151(1) &12 10 32.66,+39 24 20.7  &ESO103-G35(2)&18 38 20.30,$-$65 25 41.0\\
Mrk766(1)  &12 18 26.48,+29 18 14.0   &3C403(2)   &19 52 15.82,+02 30 24.3\\
NGC4388(2) &12 25 46.93,+12 39 43.3   &           &  \\
\hline
\hline
\end{tabular}                                                                                                                                                                               
\end{center}

%\label{table1:table1}                                                                                                                                                                    
\small                                                                                                                                                                                     
\end{table*}                                                                                                                                                                     

\begin{table*}
\begin{center}
% \begin{minipage}{140mm}
%\twocolumn
\centerline{Table 3A: Maser galaxies in the Diamond-Stanic et al. (2009) sample}
 \begin{tabular}{l r r r}
\hline\hline
Name (class)&RA(J2000), Dec(J2000) & Name(class)& RA(J2000), Dec(J2000) \\
\hline
NGC1068(2)& 02 42 40.70,$-$00 00 48.0 &NGC4258(2) &12 18 57.62,+47 18 14.0 \\
NGC1386(2)& 03 36 46.24,$-$35 59 57.0 &NGC4388(2) &12 25 46.93,+12 39 43.3 \\
NGC2273(2)& 06 50 08.67,+60 50 44.8   &NGC4945(2) &13 05 27.28,$-$49 28 04.4 \\
NGC2639(2)& 08 43 38.09,+50 12 19.9   &NGC5128(2) &13 25 27.61,$-$43 01 08.8\\
NGC3081(2)& 09 59 29.54,$-$22 49 34.6 &Circinus(2)&14 13 08.90,$-$65 20 27.0\\
NGC3079(2)& 10 01 57.80,+55 40 47.2   &NGC5506(2) &14 13 14.87,$-$03 12 27.0  \\
IC2560(2) & 10 16 18.71,$-$33 33 49.7 &NGC5643(2) &14 32 40.70,$-$44 10 28.0\\
NGC3735(2)& 11 35 57.33,+70 32 08.1   &NGC5728(2) &14 42 23.90,$-$17 15 11.0\\ 
NGC3783(1)& 11 39 01.78,$-$37 44 01.7 &NGC6300(2) &17 16 59.47,$-$62 49 14.0  \\
NGC4051(1)& 12 03 09.62,+44 31 52.8   &NGC7479(2) &23 04 56.67,+12 19 22.4  \\
NGC4151(1)& 12 10 32.66,+39 24 20.7   &           & \\       
\hline
\hline
\end{tabular}                                                                                                                                                                               
\end{center}                                                                                                                                                                               
%\label{table1:table1}                                                                                                                                                                    
\small                                                                                                                                                                                     
\end{table*}

\setcounter{table}{0}   
\begin{landscape}
\begin{table}
%\smallsize
%\scriptsize
\centering
\caption{\textbf{Table 1A: Galaxies detected by \emph{INTEGRAL}/IBIS.}}
\label{tab1}
\begin{tabular}{l l l  c l l l c c r r r}

\hline\hline

 Name                          &      RA       &      Dec        &  $z$       &   Class      & F$_{HX}^{\dagger}$  & F$_{X}^{\dagger}$ & Log N$_H$  & Notes &  Rms$^{\rm *}$  &  $L_{\rm H2O}^{\rm **}$ & Ref.$^{\rm ***}$ \\

\hline

IGR J00040+7020                & 00 04 01.92   & $+$70 19 18.5   &   0.096    &    Sy2        &   1.47    &    0.35     &  22.52    &          &     &         &    \\

IGR J00158+5605                & 00 15 54.19   & $+$56 02 57.5   &   0.169    &    Sy1.5      &   $<$0.66 &   0.31      &  21.50    &          &     &         &     \\

IGR J00256+6821                & 00 25 32.50   & $+$68 21 44.0   &   0.012    &    Sy2        &   1.47    &   0.05      &  23.60    &          &     &         &     \\

{\bf IGR J00333+6122}          & 00 33 18.34   & $+$61 27 43.3   &   0.105    &    Sy1.5      &   1.38    &   0.68      &  21.93    &    O     &  2  & $<$14.0 & this work \\

SWIFT J0034.5-7904	       & 00 34 16.7    & $-$79 05 20     & 0.07400    &    Sy1        &   1.09    &   6.33      &  23.49    &          &     &         &      \\

{\bf 1ES 0033+595}             & 00 35 52.60   & $+$59 50 05.0   &   0.086    &    BL Lac     &   1.83    &    5.90     &  21.55    &    O     &  2  & $<$9.4  & this work \\

IGR J00465-4005                & 00 46 20.68   & $-$40 05 49.1   &   0.201    &    Sy2        &   3.56    &  0.12       &  23.38    &          &     &         &    \\
\hline																				       
\hline																				       
\end{tabular}																			       
\end{table}																			       
\end{landscape} 		
\newpage

\setcounter{table}{0}   
\begin{landscape}
\begin{table}
%\smallsize
%\scriptsize
\centering
\caption{\textbf{continue}}
\label{tab1}
\begin{tabular}{l l l  c l l l c c r r r}

\hline\hline

 Name                   &      RA       & Dec             &  $z$        &   Class   & F$_{HX}^{\dagger}$  & F$_{X}^{\dagger}$ & Log N$_H$  &  Notes  & Rms$^{\rm *}$ & $L_{\rm H2O}^{\rm **}$ & Ref.$^{\rm **}$ \\

\hline

1H 0323+342                    & 03 24 41.16   & $+$34 10 45.8   &   0.061    &    NLSy1       &   3.96    &   0.64      &  21.16    &           &     &         &      \\

IGR J03249+4041-SW             & 03 25 13.20   & $+$40 41 55.0   &  0.0477    &    Sy2        &  1.35     &  0.12       & $>$21.18  &           &     &         &      \\

IGR J03249+4041-NE             & 03 25 12.20   & $+$40 42 02.0   &  0.0475    &    Sy2        &  1.35    &  0.09        & 22.48     &           &     &         &       \\

IGR J03334+3718                & 03 33 18.79   & $+$37 18 11.1   &  0.05583   &    Sy1.5      &   2.61   &  0.61        &   21.15   &           &     &         &      \\

NGC 1365                       & 03 33 36.31   & $-$36 08 27.8   &   0.0054   &    Sy1.9      &   4.10   &   1.30       &  24.65    &     O     &  3  &  $<$0.1 & MCP \\

NRAO 140                       & 03 36 30.10   & $+$32 18 29.0   &  1.2580    &    QSO/bl &   2.46   &  0.72        &  21.10     &          &     &          &      \\

ESO 548-G81                    & 03 42 03.7    & $-$21 14 40.0   &   0.0145   &    Sy1        &  4.45    &  1.35        &  20.36     &  O       &  5  &  $<$0.7 & MCP \\
\hline																				       
\hline																				       
\end{tabular}																			       
\end{table}																			       
\end{landscape}

\setcounter{table}{0}   
\begin{landscape}
\begin{table}
%\smallsize
%\scriptsize
\centering
\caption{\textbf{continue}}
\label{tab1}
\begin{tabular}{l l l  c l l l c c r r r}

  \hline\hline

 Name              &      RA     & Dec          &  $z$  &   Class   & F$_{HX}^{\dagger}$  & F$_{X}^{\dagger}$ & Log N$_H$  &  Notes  & Rms$^{\rm *}$ & $L_{\rm H2O}^{\rm **}$ & Ref.$^{\rm ***}$ \\

\hline

{\bf Mrk 3}             & 06 15 36.31   & $+$71 02 14.9   &   0.0135     &    Sy2       &   9.53   &   0.65  &  24.00     &    M(J?,b)    &         & 5$\pm$5 & MCP \\

IGR J06233-6436         & 06 23 07.70   &  $-$64 36 20.0  & 0.12889      &   Sy1        &  1.56    &  0.46   &  20.59     &         &         &         &   \\

IGR J06239-6052         & 06 23 45.61   & $-$60 58 45.4   &   0.0405     &    Sy2       &   $<$0.98  &   0.32  &  23.28   &         &         &         &     \\

SWIFT J0623.8-3215	& 06 23 46.4    & $-$32 13 00     &  0.02243     &    Sy2	&  1.94    & 1.16    &  23.91     &    O    &  4      & $<$1.3  & MCP \\

PKS 0637-752	        & 06 35 46.5    & $-$75 16 17 	  &  0.65300     &  QSO/bl  &  1.66    & 4.72    &    21.10        &         &         &         &    \\

SWIFT J0640.4-2554      & 06 40 11.69   & $-$25 53 43.4   &   0.0248     &    Sy1.2     &   3.83   &   1.88  &  21.38     &    O    & 13 (1.3) &  $<$21.9 & K06 \\

IGR J06415+3251         & 06 41 23.00   & $+$32 55 39.0   &  0.01719     &    Sy2       &   4.74   &   0.33  &  23.20     &    O    & 3      &   $<$0.6 & MCP \\
\hline																				       
\hline																				       
\end{tabular}																			       
\end{table}																			       
\end{landscape}

\setcounter{table}{0}   
\begin{landscape}
\begin{table}
%\smallsize
%\scriptsize
\centering
\caption{\textbf{continue}}
\label{tab1}
\begin{tabular}{l l l  c l l l c c r r r}

\hline\hline

 Name              &      RA     & Dec          &  $z$  &   Class   & F$_{HX}^{\dagger}$  & F$_{X}^{\dagger}$ & Log N$_H$  &  Notes  & Rms$^{\rm *}$ & $L_{\rm H2O}^{\rm **}$ & Ref.$^{\rm ***}$ \\

\hline

{\bf SWIFT J1009.3-4250}  & 10 09 48.22   & $-$42 48 40.4   &  0.033   &    Sy2         &   2.81     &   0.31    &  23.41       &     O     & 10    &  $<$6.9 & this work \\

IGR J10147-6354           & 10 14 15.55   & $-$63 51 50.1   &  0.202   &    Sy1.2       &   $<$1.07  &   0.21    &  22.30       &           &       &       &   \\

NGC 3227                  & 10 23 30.61   & $+$19 51 53.8   &  0.0038  &    Sy1.5       &  11.30     &   0.81    &  22.83       &     O     &  3    &  $<$0.03 & MCP \\

{\bf NGC 3281}            & 10 31 52.06   & $-$34 51 13.3   &  0.0107  &    Sy2         &   5.11     &    0.29   &  24.18       &     O     &  3    &  $<$0.2 & MCP \\

{\bf SWIFT J1038.8-4942}  & 10 38 45.20   & $-$49 46 53.2   &  0.06    &    Sy1.5       &   2.00     &   1.45    &  21.79       &           &       &       &   \\

{\bf IGR J10404-4625}     & 10 40 22.55   & $-$46 25 25.7   &  0.024   &    Sy2         &   3.36     &   1.20    &  22.61       &           &       &       &   \\

MCG+04-26-006             & 10 46 42.67   & $+$25 55 52.5   &  0.02    &    L       &   3.19     &   0.22    &  23.09       &     O     &   4   &  $<$1.0 & MCP \\
\hline																				       
\hline																				       
\end{tabular}																			       
\end{table}																			       
\end{landscape} 

\setcounter{table}{0}   
\begin{landscape}
\begin{table}
%\smallsize
%\scriptsize
\centering
\caption{\textbf{continue}}
\label{tab1}
\begin{tabular}{l l l  c l l l c c r r r}

\hline\hline

 Name              &      RA     & Dec          &  $z$  &   Class   & F$_{HX}^{\dagger}$  & F$_{X}^{\dagger}$ & Log N$_H$  &  Notes  & Rms$^{\rm *}$ & $L_{\rm H2O}^{\rm **}$ & Ref.$^{\rm ***}$ \\

\hline

{\bf IGR J12415-5750}        & 12 41 25.74   & $-$57 50 03.5    &  0.0242  &    Sy1.5       &   2.04    &   0.77   &  21.48       &    O    &  10 (1.3) & $<$16.1 & K06 \\

IGR J1248.2-5828             & 12 47 57.84   & $-$58 30 00.2    &  0.028   &    Sy1.9       &   1.02    &   0.40   &  22.30       &         &          &         &   \\

NGC 4748                     & 12 52 12.40   & $-$13 24 53.0    &  0.0146  &    NLSy1        &   1.27    &    0.34  &  20.56       &    O    &    3     & $<$ 0.4 & MCP \\

ESO 323-32                   & 12 53 20.19   & $-$41 38 07.5    &  0.016   &    Sy2         &  1.81     &   0.40   &    25.00     &    O    &    13 (1.3) & $<$0.1 & K06 \\

{\bf 3C 279}                 & 12 56 11.17   & $-$05 47 21.5    &  0.5362  &    QSO/bl  &   2.15    &    0.60  & 20.35        &          &         &          & \\

Mrk 783                      & 13 02 58.84   & $+$16 24 27.5    &  0.0672  &    NLSy1        &   2.21    &   0.65   &  21.18       &    O     &    2    & $<$5.7 & T11 \\

IGR J13038+5348              & 13 03 59.43   & $+$53 47 30.1    &  0.03    &    Sy1.2       &   2.74    &   1.59   &  20.22       &    O     &    4    &  $<$2.3 & MCP \\
\hline																				       
\hline																				       
\end{tabular}																			       
\end{table}																			       
\end{landscape}

\setcounter{table}{0}   
\begin{landscape}
\begin{table}
%\smallsize
%\scriptsize
\centering
\caption{\textbf{continue}}
\label{tab1}
\begin{tabular}{l l l  c l l l c c r r r}

\hline\hline

 Name              &      RA     & Dec          &  $z$  &   Class   & F$_{HX}^{\dagger}$  & F$_{X}^{\dagger}$ & Log N$_H$  &  Notes  & Rms$^{\rm *}$ & $L_{\rm H2O}^{\rm **}$ & Ref.$^{\rm ***}$ \\

\hline

H 1426+428                   & 14 28 32.57   & $+$42 40 24.8    &  0.1291  &    BL Lac      &   1.79    &   3.38   &  21.01       &          &         &         &   \\

IGR J14301-4158              & 14 30 12.17   & $-$41 58 31.4    &  0.0039  &    Sy2         &   0.92    &   0.27   & 22.08        &          &         &         &   \\

NGC 5643                     & 14 32 40.70   & $-$44 10 28.0    &  0.0040  &    Sy2         &   1.10    &   0.084  &  23.85       &    M(?)     &         & 13$\pm$5 & MCP  \\

NGC 5674	             & 14 33 52.2    & $+$05 27 30 	& 0.02493  &   Sy1.9	    &   1.81    &   4.85   &  23.05       &    O     &    4    & $<$1.6  & MCP \\

SWIFT J1436.8-1615	     & 14 36 49.6    & $-$16 13 41 	& 0.14454  &   Sy1/QSO	    &   2.00    &   7.19   &   20.88           &          &         &         &  	\\

NGC 5728                     & 14 42 23.90   & $-$17 15 11.0    &  0.0093  &    Sy2         &   5.38    &   0.15   &  24.14       &    M(D,c)     &         & 100$\pm$15 & MCP  \\

IGR J14471-6414              & 14 46 28.26   & $-$64 16 24.3    &  0.053   &    Sy1.2       &   1.11    &   0.48   &  21.60       &          &         &         &  \\
\hline																				       
\hline																				       
\end{tabular}																			       
\end{table}																			       
\end{landscape}

\setcounter{table}{0}   
\begin{landscape}
\begin{table}
%\smallsize
%\scriptsize
\centering
\caption{\textbf{continue}}
\label{tab1}
\begin{tabular}{l l l  c l l l c c r r r}

\hline\hline

 Name              &      RA     & Dec          &  $z$  &   Class   & F$_{HX}^{\dagger}$  & F$_{X}^{\dagger}$ & Log N$_H$  &  Notes  & Rms$^{\rm *}$ & $L_{\rm H2O}^{\rm **}$ & Ref.$^{\rm ***}$ \\

\hline

NGC 6221                  & 16 52 46.32   & $-$59 13 00.8    &  0.005   &    Sy2        &   1.06       &   1.41   &  22.04      &    O     & 11 (1.3) & $<$0.8 & K06 \\

ESO 138-1                 & 16 51 20.21   & $-$59 14 04.2    &  0.0091  &    Sy2        &   1.06       &   0.180  &  24.18      &     O    & 13 (1.3) & $<$3.0 & K06 \\

NGC 6240                  & 16 52 58.97   & $+$02 24 01.7    &  0.0245  &    Sy2/L      &   5.42       &    0.20  &  24.30      &    M(D?,j)     &     &  40$\pm$20 & MCP \\

Mrk 501                   & 16 53 52.22   & $+$39 45 36.6    &  0.0337  &    BL Lac     &   4.32       & 16.7     & 20.07        &   O     &  4  &   $<$2.9 & MCP \\

{\bf IGR J16558-5203}     & 16 56 05.62   & $-$52 03 40.9    &  0.054   &    Sy1.2      &   2.89       &   1.75   &  23.27       &         &     &     &     \\

{\bf SWIFT J1656.3-3302}  & 16 56 16.85   & $-$33 02 11.1    &  2.4     &    QSO/bl &  3.04        &  0.44    &  21.34       &         &     &     &      \\

IGR J17009+3559           & 17 00 53.00   & $+$35 59 56.2    & 0.113    & XBONG         &  2.13        &  0.11    & 23.7         &         &     &     &     \\
\hline																				       
\hline																				       
\end{tabular}																			       
\end{table}																			       
\end{landscape}

\setcounter{table}{0}   
\begin{landscape}
\begin{table}
%\smallsize
%\scriptsize
\centering
\caption{\textbf{continue}}
\label{tab1}
\begin{tabular}{l l l  c l l l c c r r r}

\hline\hline

 Name              &      RA     & Dec          &  $z$  &   Class   & F$_{HX}^{\dagger}$  & F$_{X}^{\dagger}$ & Log N$_H$  &  Notes  & Rms$^{\rm *}$ & $L_{\rm H2O}^{\rm **}$ & Ref.$^{\rm ***}$ \\

\hline

SWIFT J1845.4+7211     & 18 45 26.2    & $+$72 11 02 	  & 0.04630   &    Sy2          &  1.00       &   4.22	  & 22.51         &    O    & 2  & $<$2.7  & MCP	\\

IGR J18470-7831        & 18 47 02.83   & $-$78 31 49.5    &  0.0743   &   Sy1           &  1.83       &   0.80    & 20.08         &    O    & 50 (0.84) & $<$489.1 & G02 \\

PBC J1850.7-1658       & 18 50 51.6    & $-$16 55 58      & -         &  AGN/RG         &  1.28       & 2.19      & 22.17         &         &    &     &    \\

IGR J18538-0102        & 18 53 48.4    & $-$01 02 30	  & 0.14500   &   Sy1	        &  1.02       &  4.0	  & 21.61         &         &    &     &    \\

ESO 25-2               & 18 54 40.39   &  $-$78 53 54.4   &  0.0292   & Sy1             &  2.15       &    0.20   &  20.92        &    O    & 16 (1.3) & $<$37.4 & K06 \\

{\bf 2E 1853.7+1534}   & 18 56 00.00   & $+$15 38 13.0    &  0.084    &    Sy1          &   2.36      &   1.22    &  21.59        &    O    &  3 & $<$13.4 & this work \\

2E 1849.2-7832         &  18 57 07.68  & $-$78 28 21.3    &  0.0420   &   Sy1           &  3.27       &   0.77    &  20.92        &    O    & 20 (1.3) & $<$96.7 & K06 \\
\hline																				       
\hline																				       
\end{tabular}																			       
\end{table}																			       
\end{landscape}

\setcounter{table}{0}   
\begin{landscape}
\begin{table}
%\smallsize
%\scriptsize
\centering
\caption{\textbf{continue}}
\label{tab1}
\begin{tabular}{l l l  c l l l c c r r r}

\hline\hline

 Name              &      RA     & Dec          &  $z$  &   Class   & F$_{HX}^{\dagger}$  & F$_{X}^{\dagger}$ & Log N$_H$  &  Notes  & Rms$^{\rm *}$ & $L_{\rm H2O}^{\rm **}$ & Ref.$^{\rm ***}$ \\

\hline

{\bf RX J2135.9+4728}    & 21 35 54.2   & $+$47 28 22.3   &  0.025   &    Sy1           &   1.79     &   0.62      &  20.30        &    O    & 3  & $<$1.2 & this work \\

1RXS J213944.3+595016    & 21 39 45.10  & $+$59 50 14.0   &  0.114   &    Sy1.5         &   1.02     &   0.59      &  21.54        &         &    &     &       \\

PKS 2149-306             & 21 51 55.52  & $-$30 27 53.7   &  2.345   &    QSO/bl    &     2.51   &    0.80     &  20.34        &         &    &     &       \\

IGR J21565+5948          & 21 56 04.20  & $+$59 56 04.5   & 0.208    &    Sy1           & 0.98       &    0.12     &   21.81       &         &    &     &       \\

SWIFT J2156.2+1724       & 21 56 15.2   & $+$17 22 53     & 0.03400  &    Sy1.8	        &   2.00     &   1.77	   &   23.33       &         &    &     &    	\\

Mrk 520                  & 22 00 41.37  & $+$10 33 08.7   &  0.0266  &    Sy1.9         &   4.13     &   0.63      &   22.38       &    O    &  2 &  $<$0.9 & MCP   \\

{\bf NGC 7172}         & 22 02 01.90   & $-$31 52 11.6   &  0.0087  &    Sy2          &   7.93    &  4.30     & 22.86        &    O     & 7 &   $<$0.3  & MCP  \\
\hline\hline

\end{tabular}																			\begin{list}{}{}
\item Sources in boldface belong to the complete AGN catalog reported in \cite{malizia09}. The format on the relevant digits in the coordinates are those originally reported \citep{malizia12, malizia16}. 
Maser types: D=disc; J=jet; O=outflow; /=or; +=plus. For possible limits on this classification see Sect. 6. References for the Maser type: a) MCP at \url{https://safe.nrao.edu/wiki/bin/view/Main/PublicWaterMaserList} (web page updated at November 10, 2018) ; b) \cite{henkel05}; c) \cite{kamali}; d) \cite{kondratko05}; e) \cite{tarchi11b}; f) \cite{humphreys}; g) \cite{ott}; h) \cite{green03b}; i) \cite{green09}; j) \cite{hagiwara15}; k) \cite{sato}; l) \cite{wang}. ${\dagger}$ 2--10 keV and 20--100 keV flux in units of 10$^{-11}$ erg cm$^{-2}$ s$^{-1}$ $^{\rm *}$ 1$\sigma$ rms in mJy, calculated over a 0.3\,km\,s$^{-1}$ wide channel, unless otherwise specified (different channel widths are reported in brackets); $^{\rm **}$ Water maser isotropic luminosities are in unit of L$_{\odot}$, where L$_{\odot}$= 3.826 10$^{33}$ erg s$^{-1}$. We calculate the upper limits for the undetected sources using the formula: $L_{\rm H2O}[\rm L_{\odot}] = 0.023 \times S [\rm Jy] \times \Delta v [\rm km\,s^{-1}] \times D^2 [\rm Mpc^2]$, where $S [\rm Jy]$ and $\Delta v [\rm km\,s^{-1}]$, are the 5$\sigma$ rms and the channel spacing, respectively; $D$ is the distance in Mpc, derived using the redshift in column~4 and $H_0$=70\,km\,s$^{-1}$\,Mpc$^{-1}$
$^{\rm ***}$ References for the rms (undetected sources) or the isotropic maser luminosity (detected sources):
B96: \citet{braatz96}; B09: \citet{bennert09} and references therein; G96: \citet{gallimore96}; G02: \citet{green02}; H84: \citet{henkel84}; K05: \citet{kameno}; K06: \citet{kondratko06a}; MCP: Megamaser Cosmology Project webpage (\url{https://safe.nrao.edu/wiki/bin/view/Main/MegamaserCosmologyProject}); O13: \citet{ott}; P03: \citet{peck}; P15: \citet{pesce}; T03: \citet{tarchi03}; T11: \citet{tarchi11b}.
$^{\rm ****}$ The maser in IGRJ16385-2057 was detected during our survey (Sect. 3.3), a detailed analysis have been reported in \citet{tarchi11b}.

\end{list}

\end{table}																			       
\end{landscape}

\section{Maser types}

Of the INTEGRAL AGN with maser detection, only six sources (IGR\,J05081+1722, NGC\,3081, NGC\,3783, NGC\,5643, NGC\,6300, and ESO\,103-G35) so far have no associated maser type. Here we discuss each source individually and provide some clues on their most likely maser type. 

IGR\,J05081+1722 is interesting from many points of view. It belongs to an IR-luminous interacting pair of galaxies that is characterised by a luminosity for the whole system (made of a combination of star formation and accretion) of log(L$_{IR}$/L$_{\odot}$)=11.2. The system is at an early stage of merger (11.3 kpc distance) and is known to host an AGN that has optically been classified as Seyfert 2 plus a normal galaxy. The AGN, which is also the component in the system that displays  water maser emission, hosts a molecular outflow and probably also a disc wind \citep{yamashita, ballo}. 
NGC\,3783 is one of the most intensively monitored Seyfert galaxy at high energies. It is known to exhibit UV absorbers plus a series of ionised X-ray absorbers that are variable in time \citep{fukumura, mehdipour}. These have generally been interpreted as associated with a strong obscuring outflow in the nuclear region.
In contrast to other sources in our sample, neither IGR\,J05081+1722 nor NGC\,3783 display strong X-ray absorption (their column densities are about 10$^{22}$ at cm$^{-2}$), suggesting that their  water maser emission might indeed be associated with a jet or outflow. Because no clear jet emission appears to be present in NGC\,3783, but only a diffuse radio emission on tens of pc--scales \citep{orienti}, the outflow remains the only option; the situation is less clear in IGR\,J05081+1722, but the outflow is a viable possibility to explain water maser emission in this source as well. \\
Single-dish maser spectra for these sources\footnote{As shown on the MCP webpage} are consistent with an outflow origin. In particular, the maser emission in IGR\,J05081+1722 is blueshifted with respect to the systemic velocity (by $\text{about}$ 100 km/s) and appears to consist of a handful of narrow components that are attached to a broader feature. The maser spectrum in NGC\,3783 is instead characterised by a group of narrow features that are close to the systemic velocity of the galaxy. Clearly, the nature of the maser cannot be uniquely assessed without high-resolution follow-up studies.

NGC\,6300 and ESO\,103-G35 are type 2 mildly absorbed AGN with column densities of about 10$^{23}$ at cm$^{-2}$. NGC\,6300 features a slightly resolved radio core at arcsecond resolution with an extension in the southern direction \citep{morganti}, and there is also evidence for a complex molecular hydrogen structure made of an edge-on outflow superimposed on a rotating disc \citep{davies}. 
ESO\,103-G35 does not show evidence for a radio jet; furthermore, an in-depth analysis of the galaxy optical properties brought no clues on the origin of water maser emission, including no evidence for outflows \citep{bennert04}. Despite this, the source has been reported in X-rays as the site of a highly ionized outflow \citep{gofford}.
As in the two previous cases, the interpretation (among others) of an outflow origin of the water maser emission does not contradict the shape of the maser spectra\footnote{As shown on the MCP webpage} for NGC\,6300 \citep{green03b} and ESO\,103-G35. In both cases, emission is detected close to (or slightly redshifted with respect to) the systemic velocity of the target, and it is comprised of a very small number of narrow features (one in the case of ESO 103-G35) that are placed above a broader component. A somewhat different spectrum of ESO\,103-G35 is shown in the detection paper \citep{braatz96}, however, where emission from a single 20-km/s wide line is shown that is clearly redshifted by $\sim$ 100 km/s with respect to the systemic velocity.

Finally, NGC\,5643 and NGC\,3081 are heavily absorbed objects (NGC\,3081 is also Compton thick). In these objects, nuclear discs are likely to be the site where maser emission develops. Recent observations with ALMA of NGC\,5643 \citep{alonso} have at the parsec scale resolved a massive rotating disc or torus of molecular gas with strong non-nuclear motion features associated with radial outflow in the disc. Interestingly, the maser emission appears to be centrally located with respect to the inner (nuclear) part of this structure, which is also tilted with respect to the larger scale disc.
The inner structure of NGC\,3081 has instead been mapped with the Gemini Multi Object Spectrograph (GMOS-IFU) \citep{schnorr} and was found to host an even more complex structure, but at the kilo-parsec scale: this includes rotation in the galaxy disc plane, a bipolar outflow from the AGN,
non-circular motions along the nuclear bar, and an interaction between the bipolar outflow and the disc gas. 
Both objects thus resemble the well-known maser sources NGC\,1068 and the Circinus galaxy, where water maser disc emission is coupled to jet or outflow maser radiation. This suggests a similar interpretation for NGC\,5643 and NGC\,3081 as well.
Because the single-dish spectrum alone\footnote{Shown on the MCP webpage}  has a relatively low signal-to-noise ratio, it is not possible to infer a secure hypothesis on the nature of the maser in  NGC\,3081. The maser in NGC\,5643 (Greenhill et al. \citeyear{green03b}, and MCP webpage) instead indicates redshifted (50--100 km/s with respect to the systemic velocity) emission that consists of a relatively broad feature with two or three peaks and might be due to a blending of features. Speculatively, the maser might then either be associated with a jet (a diffuse radio jet on either side of the nucleus is indeed visible in a high-sensitivity image obtained with the Very Large Array, Leipski et al. \citeyear{leipski}) or with a rotating structure of which we only see the redshifted lines or the systemic lines when a large uncertainty in the reported target recessional velocity is accounted for. 

As mentioned before, however, confident clues on the association of the maser emission with the AGN activity and on the maser nature of all six objects must await follow-up (interferometric) studies. In particular, all these maser sources, with the exception of ESO\,103-G35 ($\sim$ 460 solar luminosities), have moderate maser isotropic luminosities: the luminosities of four targets are only slightly higher than the paradigmatic threshold that separates kilo- and megamasers (35, 20, 17, and 13 solar luminosities for IGR\,J05081+1722, NGC\,3783, NGC\,3081, and NGC\,5643, respectively), and one object, NGC\,6300 ($\sim$ 3.5 solar luminosities), is below that threshold. While this is still consistent with an outflow-related origin of the maser emission (Tarchi et al. \citeyear{tarchi11a} discuss an analogous origin for the water maser detected in a sample of NLSy1 that has similar luminosities to ours), an association with star formation activity for these masers, especially for that in NGC\,6300, cannot be ruled out a priori.\\

\end{document}